\documentclass[12pt,preprint]{aastex}
\slugcomment{Accepted  for publication  
in {\it The Astrophysical Journal}}  
\def\lax {\ifmmode{_<\atop^{\sim}}\else{${_<\atop^{\sim}}$}\fi}  
\def\gax {\ifmmode{_>\atop^{\sim}}\else{${_>\atop^{\sim}}$}\fi}  
\def\gtorder{\mathrel{\raise.3ex\hbox{$>$}\mkern-14mu
             \lower0.6ex\hbox{$\sim$}}}

\def\cm2{cm$^{-2}$}
\def\s1{s$^{-1}$}

\def\sax{{\it BeppoSAX}}

\begin{document}


\title{{\it BEPPOSAX} and {\it RXTE} SPECTRAL STUDY OF THE LOW-MASS X-RAY BINARY 4U~1705-44. SPECTRAL HARDENING DURING the BANANA BRANCH
}




\author{  Elena Seifina\altaffilmark{1}, Lev Titarchuk\altaffilmark{2},  Chris Shrader\altaffilmark{3,4} \& Nikolai Shaposhnikov \altaffilmark{3,5} }
\altaffiltext{1}{Moscow State University/Sternberg Astronomical Institute, Universitetsky 
Prospect 13, Moscow, 119992, Russia; seif@sai.msu.ru}
\altaffiltext{2}{Dipartimento di Fisica, Universit\`a di Ferrara, Via Saragat 1, I-44122 Ferrara, Italy, email:titarchuk@fe.infn.it; 
  Goddard Space Flight Center, NASA,  code 663, Greenbelt  
MD 20770, USA; email:lev@milkyway.gsfc.nasa.gov, USA}
\altaffiltext{3}{NASA Goddard Space Flight Center, NASA, Astrophysics Science Division, Code 661, Greenbelt, MD 20771, USA; Chris.R.Shrader@nasa.gov}
\altaffiltext{4}{Universities Space Research Association, 10211 Wincopin Cir, Suite 500, Columbia, MD 21044, USA}
\altaffiltext{5}{CRESST/University of Maryland, Department of Astronomy, College Park MD 20742, Goddard Space Flight Center, NASA,  code 663, Greenbelt  
MD 20771,  USA: email:nikolai.v.shaposhnikov@nasa.gov}

\begin{abstract}
We  analyze  the X-ray  spectra  of the   atoll 4U~1705-44 when the source undergoes 
the   island$-$banana state transition. 
We use  the {\it Rossi} X-ray Timing Explorer ({\it RXTE})  and {\it BeppoSAX} observations for this analysis.  We demonstrate  that the 
broad-band energy 
spectral distributions for all 
evolutinary  states 
can be 
fitted by  a 
model, consisting 
{two Comptonized 
components. One arises from the seed  photons coming from a neutron star  (NS) atmosphere at  a temperature
$kT_{s1}\lax$1.5 keV (herein {Comptb1})  and a second resulting from 
  the seed photons of $T_{s2}\sim$1.1$-$1.3 keV coming from the disk
(herein {Comptb2})}.  We found that we  needed to add a low-temperature  blackbody and an iron-line ({Gaussian}) component to the model in order to obtain high$-$quality fits.  
The data analysis using this model indicates  that 
the {power-law photon} index $\Gamma_{1}$ of our model 
is always about 2, independently of the spectral state.
Another parameter, $\Gamma_{2}$ 
demonstrates a two-phase behavior depending  on the spectral state. 
$\Gamma_{2}$ is quasi-constant at $\Gamma_{2}\sim 2$ when the electron temperature 
$kT^{(2)}_e<80$ keV and 
$\Gamma_{2}$ is  less than 2,  in the range of  
$1.3<\Gamma_{2}<2$, when $kT^{(2)}_e>80$ keV.
This phase is similar to that was previously found in 
the {\it Z}-source  Sco X-1. We interpret the decreasing index  phase using a model
in which a  super-Eddington  radiation pressure from the neutron star   causes  an expansion of the Compton cloud similar to that found previously in Sco~X-1 during the Flaring branch.
\end{abstract}

\keywords{accretion, accretion disks---black hole physics---X-ray:binaries---stars: individual (4U 1705-44)---radiation mechanisms: nonthermal}

\section{Introduction}

Low mass X-ray binaries (LMXBs), which contain 
a neutron star (NS),  have been observationally classified 
as {atoll}-type and {\it Z}-type XBs (Hasinger \& van der Klis, 1989). 
The distinct observational characteristics of  these two groups are not fully understood.
This classification was initially based on the patterns  which could be traced out using the 
Color Color Diagram (CCD). The CCD of  {\it Z-}sources shows
 three branches 
[horizontal, normal and flaring ones
(HB, NB and FB respectively) or 
so called a {\it Z-}pattern, 
while the CCD of atolls displays  two branches. 
These two branches of {atolls}  are known in the literature  
as the island and the banana states. However later  it has been also found 
that these two classes exhibit quite different spectral and timing properties.

The main difference between 
these two groups is probably  
related to  their different luminosity ranges. 
 Most  atolls vary 
in the range from  0.01 to 0.5 of the Eddington limit $L_{Edd}$, 
while {\it Z-}sources irradiate  
close to  $L_{Edd}$. Their luminosity is  in the  (0.5 - 1)~$L_{Edd}$ range. In this context, 
Lin et al. (2009) and 
Homan et al. (2010)   made an interesting suggestion for the {\it Z-}track sources  
using  {\it RXTE} observations of the transient neutron  source NS XTE~J1701-462.  The source   behaves  
 as a {\it Z-}source when its luminosity was  close to $L_{Edd}$, and when the source fades, it shows all properties  of an  {\it atoll}.   
However, the observations of  XTE~J1701-462 for the last   
seven years no longer demonstrate  this type of behavior.

Another  example of the hybrid atoll and {\it Z-}source 
properties is revealed in  
a unique source 4U~1705-44. It shows  
the spectral soft-hard state evolutions 
on the time scales of a month with well-determined  recurrence. 
 Also Langmeier et al. (1987); Sztajno et al. (1985) and Di Salvo et al., 2015)  
found 
type I X-ray bursts  and  Ford et al. (1998) discovered  kilohertz quasi-periodic oscillations. 
While 
4U~1705-44 is identified
as an atoll, 
it 
also has  
properties of a {\it Z}-source (Barret \& Olive, 2002, hereafter BO02). 
%
The power density spectra (PDS) of 4U~1705-44  resemble   those of {\it Z-}sources 
when it 
is on its diagonal branch (similar to the $normal$ 
branches of {\it Z-}sources). In this case 
its PDS resembles those
found on the top branch of the {\it Z}-track 
(see the {\it RXTE} data analysis in   BO02).




 Usually a sum of  a soft  {\it blackbody} component, presumably forming in the accretion disk (AD),  
 and  a hard X-ray component  originating in  a corona [a transition layer (TL), located between the NS and the AD] well describes X-ray spectra of NS LMXBs .
In addition, a broad iron line (most likely the Fe K emission) is often revealed. 
BO02 analyzed  the {\it RXTE} PCA data 
of 4U~1705-44  focusing on the spectral-timing 
behavior. They used this model and included a  6.4 keV
iron line component to show that the spectral transitions of 4U
1705-44 were followed by changes of the plasma temperature of the
Comptonizing cloud. However, the BO02 model did not separate  the contribution of the Comptonized  
disk photons from  that of the NS photons in the resulting spectrum.
Paizis et al. (2006)  and Fiocchi et al. (2007)  presented  the average INTEGRAL IBIS spectrum of 4U~1705-44 which indicated  the existence  of a high-energy tail in the spectrum.

Recently   Seifina et al (2013), hereafter STF13, applied a so-called {two component Comptonization} model to the spectral data of  
GX~340+0  and  Titarchuk et al. (2014, hereafter TSS14),  analyzed 
the data of 
Sco~X-1 using the same  model.  Note that GX~340+0  and Sco~X-1 are {\it Z-}sources. 
This self-consistent model  allows one to separate Comptonized emission caused by upscatering  of  the NS surface photons  in a Compton cloud  from  that caused by  the seed photons of the accretion disk. 
STF13 demonstrated  that the Comptonized emission of the NS photons dominates during all spectral states for  GX~340+0 
and Sco~X-1, while  that related to the Comptonized disk photons    formed the hard energy tail of X-ray spectrum. 
Previously, Farinelli et al. (2009)  successfully implemented  the same kind of model
 in an  analysis of the  {\it Beppo}SAX observations of {\it Z-}source Cyg X-2  to  describe the spectral evolution  of the  source  from  the HB to the NB.


{The broad-band ($0.1 - 200$ keV) spectrum of 4U 1705-44 was analyzed by
Egron et al. (2013), who used the combined data of {\it Beppo}SAX,  {\it XMM-Newton}, 
and {\it RXTE}. The authors  fitted the spectrum  by a combination of a BB plus a
thermal Comptonized model ({\tt compTT} or {\tt nthcomp}, in the XSPEC
package), modified by the 
photoelectric absorption ({\it phabs}) at low energies. Their fit was modified  by the addition of a 
by a broad  {\tt Gaussian} line 
and also
by inclusion of the reflection components {\tt pexriv} or {\tt reflionx}.
Their results are in agreement with those reported by Piraino et al. (2007) and
Di Salvo et al. (2009) for the soft state of 4U 1705-44. In particular,
these authors found that in the soft state,  
$N_H$ varies from  $1.96\times 10^{22}$ cm$^{-2}$ to $3.64 \times 10^{22}$ cm$^{-2}$  and the index $\Gamma$ changes from 2.2 to 2.6,
the BB temperature is about 0.5 keV, the plasma and seed photon
temperatures of the Comptonized component are 3.5 keV and 1.2 keV,
respectively. The centroid energies of four {\it line} components are
at 2.6 keV, $3.31 \pm 0.03$ keV, $3.88 \pm 0.03$ keV and $6.63 \pm 0.01$
keV, respectively. Egron et al. (2013) used a similar model to fit the
hard state spectra of 4U~1705-44. They found that in the hard state 
$N_H$ is about $2.0 \times 10^{22}$
cm$^{-2}$, the  index $\Gamma$ changes from 1.8 to 2.0, the
BB temperature ranges from 0.24 keV to 0.58 keV, and the electron and
seed photon temperatures of the Comptonized component are  about $21$ 
keV and 0.7 keV, respectively.
 The inclination of the system,
estimated using the {\tt diskline} or the reflection component, is in the
interval of $30^\circ - 40^\circ$ [see Egron et al. (2013); D'Ai et al.
(2010); Di Salvo et al. (2009) and Di Salvo et al. (2015)]. Note that
XMM-{\it Newton} data analysis done by Egron et al. (2013) provides a value
$N_H$ around $2 \times 10^{22}$ cm$^{-2}$ for both of the hard
and soft states. They also find that the $N_H$ value derived from the
non-simultaneous {\it RXTE} data is different, $N_H = (3.64 \pm 0.02) \times
10^{22}$ cm$^{-2}$.
}

Recently, \cite{Church14}  
present  the relation between atoll and {\it Z-}sources in the frame of a unified model for LMXBs. 
Based on the {\it RXTE} and $Beppo$SAX data analysis 
for a number of {\it Z} and atoll sources (4U~1705-44 among them) they used the so-called   ADC (the Comptonizing accretion 
disk corona) model to show how the ISs and BSs  in atolls relate 
the states seen
in  {\it Z} sources. Particularly,  they demonstrated a common feature of  {\it Z} and atoll behavior of the electron temperature $T_e$ being  high in  the hard states  and then it sharply dropping at the ``critical'' luminosity of about  
$2\times 10^{37}$ erg s$^{-1}$. But $T_e$ is   low (at about $T_e\sim T_{BB}$) when the luminosity increases during the soft states. 

\cite{Church14} also claim that the IS of atoll sources is characterized 
by a high $T_e$   based on the ADC model but  the IS is absent   in {\it  Z-}sources because  their  luminosities are   greater than a ``critical'' value.  Also in {\it Z-}sources there is 
the flaring Branch (FB) which is associated 
with unstable nuclear burning on the NS
at high mass accretion rate $\dot M$. This unstable burning regime is not
 pronounced at low $\dot M$  and thus this branch is not observed  in atoll  sources (except in the GX atolls). 
However,  \cite{Church14} are not able to explain the presence of  a high energy tail seen up to 200 keV   in {\it Z-}sources (D'Ai et al. 2007)  
and in the soft state of atoll-sources. In particular, Di Salvo et al. (2000) discovered a state-dependent hard energy tail in {\it Z-}source GX~17+2. For example, the prominent {tail} in Sco X-1 was  first revealed by D'Amico et al. (2001) with HEXTE/{\it RXTE} 
and Di Salvo et al. (2006) using  $BeppoSAX$. The first {\it hard tail} in $atoll$-type source 
was discovered by Piraino et al. (2007).
Thus, it is necessary to take into account these high-energy tails in analyzing the spectra 
of atoll and {\it Z} sources.


 TSS14 studied the spectral-timing property relations   observed in  Sco~X-1 using the {\it RXTE} observations. 
They found that the Sco~X-1  energy spectra    [for  (3$-$200) keV range]  
during all {\it Z}-state transitions could be adequately reproduced by a {double-Comptonization}  model. This model allows us to separate two contributions to the resulting spectrum. One spectral component is  presumably associated with  the Comptonization of the NS seed  photons and another one is related to the Comptonization of the  seed photons from the disk.
Using this model, TSS14 reveal
a  unique stability of $\Gamma_1$ and $\Gamma_2$ 
 which values are around 2 during the {\it Z-}states except of the FB. 
TSS14 explain  the detected stability of these photon  indices using 
the model in which 
the  disk photon flux is negligibly small  with respect to the energy release  in the TL. 
However, over the FB  TSS14 detect {the decreasing index phase},
which they interpret using a  model in which the accreted matter deposits its gravitational energy 
only in some external part of the  TL.  
In this case the TL innermost part is cooler  than the outer part (mostly owing to illumination by the NS soft photons), and thus the resulting radiative luminosity $L_r$ achieves   the Eddington one there.   As a result,  the accretion flow is stopped by $L_{r, \rm Edd}$. 

In this paper  we   analyze  the  {\it BeppoSAX}  observations of 4U 1705-44 that took place
during August and October of 2000 
{
[see \cite{Fiocchi07}, Piraino et al. (2007) , \cite{egron13}].   
}
 We also analyze   the {\it RXTE} observations that were made during  1997 -- 2000 
{
[see \citet{barret02}, \citet{muno02} and \cite{egron13}] 
}
In \S 2 we present  the details of our data selection procedure and show  
an  observation list that  we selected  for  our  analysis.   
In \S 3 we give  the spectral analysis details. 
 We interpret the spectral properties and their  evolution 
observed during the different spectral states in \S 4.  
Our results  are discussed and our conclusions are presented  
in  \S 5. 

\section{Data Selection \label{data}}

In 2000 August   and October  $BeppoSAX$ observed 4U 1705-44  (see details in Table 1).  
A combination  of the data from  three {\it BeppoSAX} instruments gives us the source energy spectra. Namely, we used 
Narrow
Field Instruments (NFIs): the Low Energy Concentrator
Spectrometer (LECS; \citet{parmar97}) for the 0.3 $-$ 4 keV range, the Medium Energy Concentrator Spectrometer
(MECS; \citet{boel97}) for the 1.8$-$10 keV range, 
the High Pressure Proportional Gas Scintillation Counter 
(Manzo et al. 1997) for the 8$-$50 keV range, 
and the Phoswich Detection System \citep{fron97} for the 15$-$150 keV range. 

For data reprocessing  we applied the SAXDAS data analysis package and we carried out
the spectral analysis in the energy band related to  each of these instruments. 
 A relative NFI normalization  was considered  as a free parameter  for  
 the model fitting. However, we fixed the MECS normalization at 1.   
Following  this procedure, we wanted to control that 
all of these  normalizations were located 
in  standard instrument  ranges
\footnote{http://heasarc.nasa.gov/docs/sax/abc/saxabc/saxabc.html}.

We also 
rebinned  the spectra 
 to obtain
significant data points.  In particular,  the LECS   spectra were rebinned  with a binning factor 
which is not constant over energy (see \S 3.1.6 of Cookbook for the {\it BeppoSAX} NFI spectral analysis). For this  rebinning we applied  
template files  in GRPPHA of  XSPEC \footnote{http://heasarc.gsfc.nasa.gov/FTP/sax/cal/responses/grouping}. 
Moreover,  
the  Phoswich Detection System spectra were rebinned  with a  
binning  factor of 2 when we  grouped  two bins together (resulting  width of the bin is 1 keV).  
We applied a systematic error of 1\%   to 
{
$BeppoSAX$ 
}
spectra.  
We list  all {\it BeppoSAX} observations used in the  presented analysis in Table 1. 

We  have also used  the {\it RXTE} data which were 
obtained from 1997 April  to 2000 February. 
In total, these data  include 86 observations sampling the different spectral states.
For data processing we utilized standard tasks of the LHEASOFT/FTOOLS
5.3 software package.
We applied  Proportional Counter Array (PCA)  Standard 2 mode data,
accumilated in the 3$-$23~keV energy range, using the most recent release of PCA response 
calibration (ftool pcarmf v11.7) for a spectral analysis. We  also applied a standard dead time correction procedure  to the data.  To construct broadband spectra, the data from HEXTE detectors were used.  A background derived from   off-source observations was subtracted from the source signal. 

We have included only data  in the 19$-$200~keV 
range.  Thus, we  accounted  for  the HEXTE response uncertainties   and 
determination of the background. 
 The data are available through the GSFC public archive 
(http://heasarc.gsfc.nasa.gov).   A list  of 
observations that covers 
the source spectral evolution 
 is present in  Table 2.

\begin{figure}[ptbptbptb]
\includegraphics[scale=0.9,angle=0]{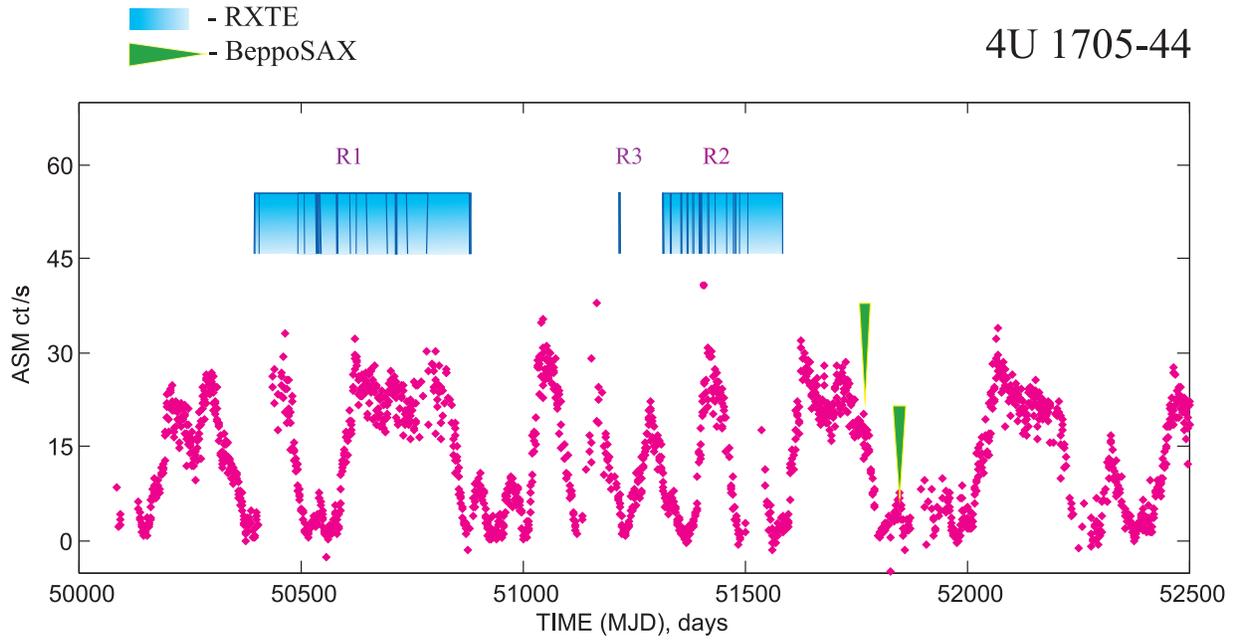}
\caption{  
 Evolution of ASM/{\it RXTE} count rate 
during  1997 -- 2000 observations of 4U~1705-44. 
 {Green} triangles show {\it BeppoSAX} NFI data, listed in Table 1, and  {blue} vertical strips (on  {\it top} of the panel) indicate a temporal distribution of the {\it RXTE} data 
of pointed observations used in our
analysis, and
{bright blue} rectangles indicate
the {\it RXTE} data sets listed in Table 2. 
}
\label{variability_97-09}
\end{figure}

In Figure  \ref{variability_97-09}  we plot the overall light curve of the All Sky Monitor (ASM; Levine et al. 1996) on board \textit{RXTE} in order to  assess the intensity of the source on long timescales. The ASM data (from 1995 October 10  to 2002 August 14 2002) indicate 
long-term variability on   the scale of 1 month from 5 ASM counts $s^{-1}$ (corresponding to the hard state) to 35 ASM counts s$^{-1}$ 
(the soft state). Note that during the hard state 
a type-I X-ray burst occured, but none occurred   
in the soft state (see D'Ai et al., 2010). 
 In fact, we have  excluded the type I bursts from the analysis.   Generally, the source showed  the hard-to-soft  transitions characterized by the different timescales of the X-ray flux variability. As seen from Fig.~\ref{variability_97-09}, we have chosen the time 
intervals with 
long (with high flux ``plateau'', $R1$ set) and with relatively shorter  timescale periods
of variability ($R2$ and $R3$). 
Thus, we have    made a  {\it RXTE}   analysis 
of the 4U~1705-44 data  for  3 yr of observations   and divided them  to  three intervals indicated by  blue rectangles in Figure~\ref{variability_97-09} ($top$).
The {\it RXTE} 
energy spectra were modeled using XSPEC astrophysical fitting software. 
We applied a systematic uncertainty  of 0.5\%    to all  analyzed 
{\it
RXTE   
}
spectra.


\section{Results \label{results}}

\subsection{CCDs 
diagrams of 4U~1705-44 \label{ccd}}

\begin{figure}[ptbptbptb]
\includegraphics[scale=0.89, angle=0]{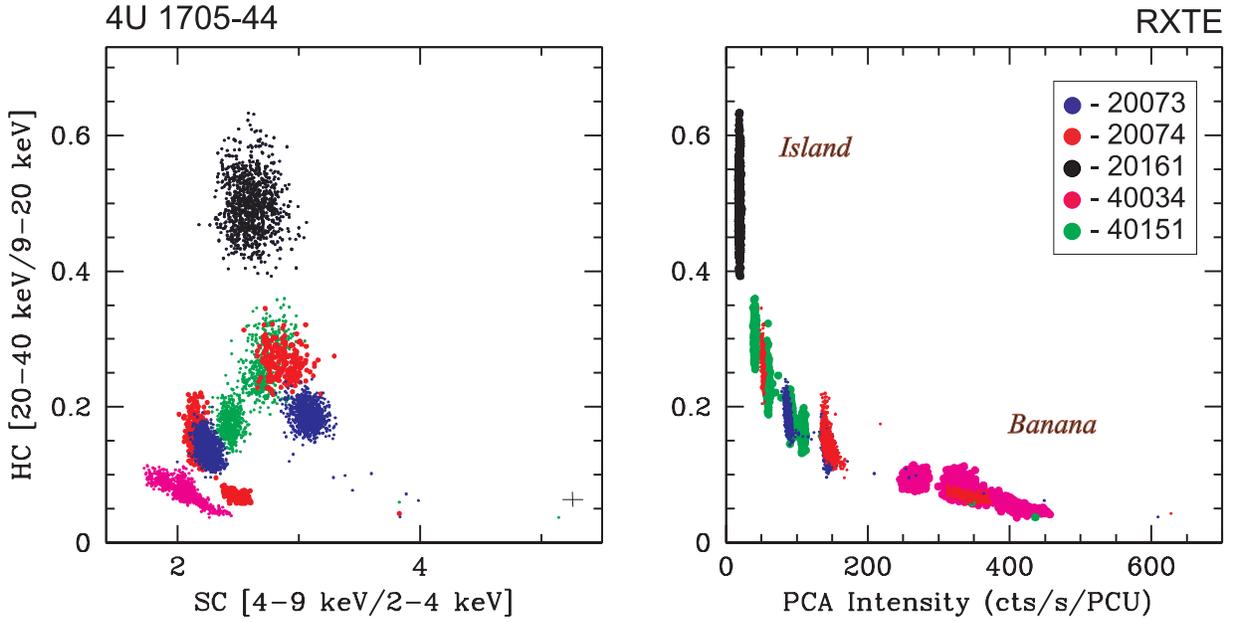}
\caption{
CDs ({left}) and HIDs ({\it right}) for all observations of 4U~1705-44 used in our analysis, with  
bin size 16 s. 
 The typical error bars for the colors are shown in the bottom right   corner of the left panel, while  errors of  the intensity are negligible.  
The sets are indicated by different colors: blue (ID 20073), red (ID 20074), black (ID 20161), crimson (ID 40034) and green (ID 40051).
}
\label{CCD}
\end{figure}

To investigate the properties of 4U~1705-44 during different spectral states, 
{we  use}   hard colors (HCs) and soft colors (SCs) and we demonstrate different configurations of 
CCDs (CCDs, HC versus SC).
In Figure~\ref{CCD} we 
combined CCDs
(left panel) and hardness-intensity diagrams (HIDs; right panel).
The axises 
of CCDs show the flux ratios, 
[20-40 keV/9-20 keV] and  [4-9 keV/2-4 keV], 
while HIDs  demonstrate  flux ratios, 
[20-40 keV/9-20 keV]  versus PCA 
count rate (2-40 keV). 
In the top right corner we indicate IDs of used observational sets. 
 One can see that our data sets cover different parts of HID and CCD from ISs to BSs.
Therefore, we select observations performed at different flux levels in the hard and soft states and proceed to use these data for a detailed spectral analysis of the properties of 4U~1705-44 during its island$-$banana state evolution. 

\begin{figure}[ptbptbptb]
\includegraphics[scale=0.91, angle=0]{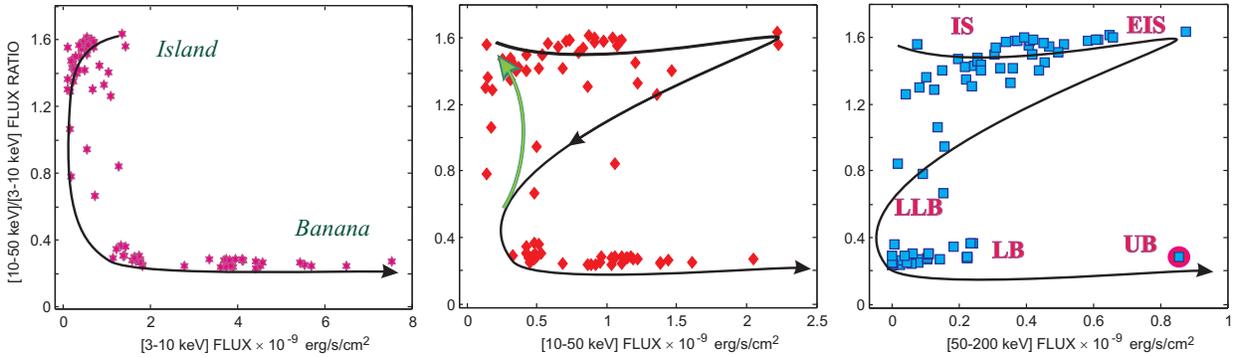}
\caption{
Left to right: 
flux ratio [10-50 keV/3-10 keV] vs flux in [3-10 keV], [10-50 keV] and [50-200 keV] 
energy ranges 
measured in units of $10^{-9}$ erg s$^{-1}$cm$^{-2}$. 
Black arrows indicate the direction of state changing from island to banana branches, 
while green arrow (middle panel) marks the direction of state changes from banana to island branch.
In the right panel we indicate spectral states of 4U~1705-44 according to van Straaten (2000), and the point 
with pink oreole which spectrum is presented in Figure~\ref{Zsp_compar_RXTE}, related to ``UB''state.
}
\label{3HID}
\end{figure}

Note that the shape of the track described above, particularly in the HID, can be affected by secular shifts,  the so-called parallel tracks
 (see, e.g. Di Salvo et al. 2003).
Therefore,  
the data of different sets were marked by different colors (see Figure~\ref{CCD}). 
While  
data sets  investigated by us are associated with different 
epochs it is clear  from this figure 
that our selected data do not show a noticeable effect of secular shifts. 
In particular, the HIDs  exhibit  plain and smooth tracks.
Furthermore, we test this effect using flux units instead of counts, 
in which case the corresponding tracks (hardness-flux diagrams, see Sect.~3.2.4 and 
Fig. \ref{3HID}) also form a plain and smooth track.



\subsection{Spectral Analysis \label{spectral analysis}}

\subsubsection{Choice of  a Spectral Model \label{model choise}}

{
Many authors have
used various models
for  fitting the 4U~1705-44 spectra.  
The hard component observed  at high luminosity  (the  high/soft state)  can be presented  
by a power-law that 
can be interpreted  as Comptonization of electrons that  have a nonthermal velocity 
distribution. 
Specifically, Egron et al. (2013) 
fit $BeppoSAX$,  $XMM-Newton$,  and {\it RXTE} spectra related to
the soft and   hard states using a model which includes a Comptonization component,  a soft BB component  and a smeared reflection component. In particular,  the latter
component is used to fit the data associated 
 with the hard state.
 A hard power-law component should  also be added in the soft state  to  account for the 
state$-$dependent hard tail (see Piraino et al. 2007) .
Our approach is to employ a single model that is capable of representing both the hard and soft spectral states of the system with a minimal number of free parameters, with the goal of probing the underlying physical process driving the system.

 To probe our modeling approach,  we proceeded initially using  a model that is a sum  of 
 a Comptonization component, soft BB,  and power-law, modified by an interstellar absorption. We also add a
Gaussian line component to take into account a broad iron line associated with the Fe K emission.
But this model [$wabs*(bbody + powerlaw + comptb+Gauss)$ provides satisfactory fits 
only for the soft state data [e.g., 40034-01-02-06 and 40034-01-02-09 spectra, $\chi^2_{red}$=1.16 and 1.34 [77 degree of freedom  (d.o.f.), 
respectively, see Table~3];  in other words,  the model gives a good 
data description   only in 
50\% of cases.  Moreover, the model requires a very large photon index (much greater than 3) for the soft$-$state spectra 
(e.g., for 40034-01-02-09 spectrum $\Gamma_{pow}$=3.97$\pm$0.05, see Table~3) and it gives unacceptable 
fits for the  hard-state data [e.g., for 21292002 spectrum of $BeppoSAX$ data $\chi^2_{red}$=3.28 (193 d.o.f.)] 
and for 40034-01-07-02G spectrum of {\it RXTE} data [$\chi^2_{red}$=1.78 (79 d.o.f.)]. 
 Thus, we  conclude that the best representation of the data requires 
a  spectral model  that can account for both   the soft and the hard components, 
where  each component can be represented by a Comptonization  model.  
We  use a sum of two Comptonization components, in which 
the first and second  components are related to  the soft and hard components 
of the spectrum respectively.
As a result, 
the best fits of the hard- and soft-state spectra have been produced  by implementation of the so-called 
{\it double Comptonization model} (using  {\tt Comptb}\footnote{{Comptb} XSPEC  model, 
see XSPEC v. 12.8.0 \\ in  http://heasarc.gsfc.nasa.gov/docs/xanadu/xspec/models/index.html} as an example of the Comptonizing model).  
 This model consists  of two Comptonized components (both due to the presence of the TL that 
upscatters seed photons of $T_{s1}\le 1.7$ keV coming from the neutron star  (first component $Comptb1$), and seed photons 
of temperature $T_{s2}\le 1.3$ keV coming from the disk  (second component $Comptb2$), a soft BB  and the iron-line (Gaussian) 
component (see  Table 3 for comparative analysis of $BeppoSAX$ and 
{\it RXTE} data). Thus, 
 we apply the {double-Comptonization model} for  fitting  all extracted spectra.
}

\begin{figure}[ptbptbptb]
\includegraphics[scale=0.89, angle=0]{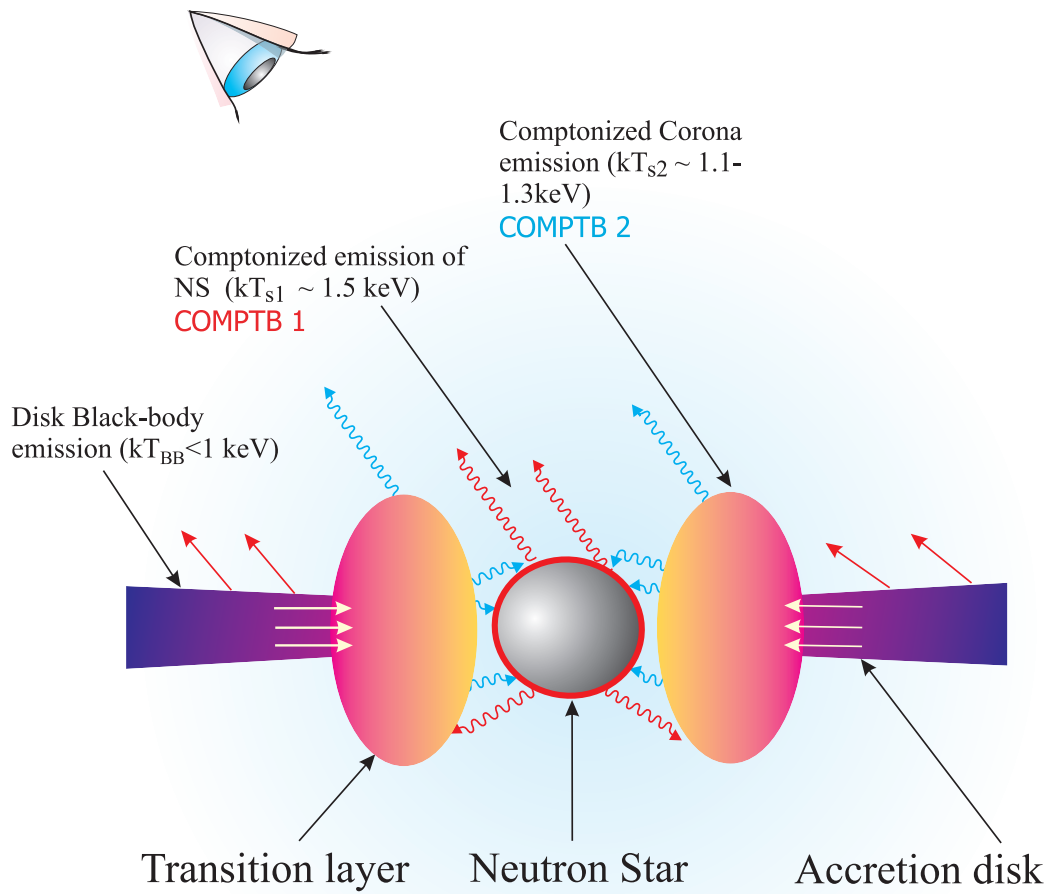}
\caption{
Suggested  geometry of 4U~1705-44.   Disk and NS soft photons are 
upscattered off   hotter plasma of the transition Layer (TL)  located 
between the accretion disk and NS surface.  Some fraction of these 
photons is seen directly by the Earth observer. Red and blue photon trajectories correspond to soft and hard (upscattered) photons respectively. 
In our model two  Comptonization components are considered.  The first one (Comptb1) is associated with  $T_{s1}\sim$1.5  keV related to the NS surface 
while {Comptb2} is related to the   {seed} (disk) photon temperature, $T_{s1}=1.1-1.3$ keV.   
}
\label{geometry}
\end{figure}

This model supports a scenario 
in which a Keplerian disk  is detached from 
the {\it Neutron star} (NS) 
by   the TL [see Titarchuk et al. (1998)].
In Figure~\ref{geometry}  this spectral model is schematically illustrated.
We assume that accretion into a NS occurs  when the plasma goes through 
an AD [standard Shakura-Sunyaev 
disk, see \citet{ss73}]
and the  TL. 
In this scenario the soft  NS   and  disk  photons   are  Comptonized by
corona (TL) electrons. As a result, the emergent  
spectrum is  produced in  the TL, where 
soft photons of the temperature $T_{s1}$ from the  NS surface and thermal disk  seed photons of the temperature $T_{s2}$  are Comptonized off the relatively hot plasma (electrons) of the TL.  The TL  forms 
two Comptonized components {\it Comptb1} and {\it Comptb2}, respectively. 
Note that the Earth observer can see directly some fraction of the NS  and disk seed  photons. 
That motivates  us to   add  a BB component with temperature $T_{BB}$ to the resulting spectrum.
In Figure~\ref{geometry} we show the soft 
(seed) and hard (upscattered) photons, as   red and blue photon trajectories respectively.

It is worth noting that in the framework of the {Comptb} the emergent spectrum is   a convolution  of the BB spectrum of the photon temperature $T_s$ and 
a normalization $N_{com}$
 with the 
upscattering   Green function.
The normalization $N_{Com}$, identical to the   XSPEC {\it bbody} model:
\begin{equation}
N_{Com}=\biggl(\frac{L}{10^{39}\mathrm{erg/s}}\biggr)\biggl(\frac{10\,\mathrm{kpc}}{D}\biggr)^2.
\label{comptb_norm}
\end{equation}  
where $L$ is the soft photon (BB) luminosty and $D$ is the distance to the source. 

To fit the data, we apply our spectral model 
$wabs*(Bbody+Comptb1+Comptb2+Gauss)$, where the fit model parameters are
the hydrogen equivalent 
absorption column 
$N_H$; the energy 
indices $\alpha_1$, $\alpha_2$
($\alpha_1=\Gamma_1-1$ and $\alpha_2=\Gamma_2-1$); 
the {\it seed} photon temperatures 
$T_{s1}$, $T_{s2}$; parameters $\log(A_1)$ and  $\log(A_2)$  related to the upscattering photon 
fractions $f_1$, $f_2$, specifically  where $f=A/(1+A)$ is  
  the relative weight of the upscattering  component; 
the plasma (electron) temperatures $T^{(1)}_e$ and  $T^{(2)}_e$, the BB normalizations 
 $N_{Com1}$ and $N_{Com2}$ of the Comptb1 and Comptb2
respectively. 
%
Finally, to fit the data in the 6$-$8 keV range,  we add a {Gaussian} component that is characterized  
by the  parameters $E_{line}$ (a centroid line energy), $\sigma_{line}$ (the line width)   
and  $N_{line}$ (the line normalization).   

In any case of $\log(A)\gg1$ a  fraction $f = A/(1 + A)$ approaches unity and thus we always fix 
values  of $\log A_{1,2}$  at 2.

\subsubsection{{\it BeppoSAX} data analysis}

\begin{figure}[ptbptbptb]
\includegraphics[scale=1.0,angle=0]{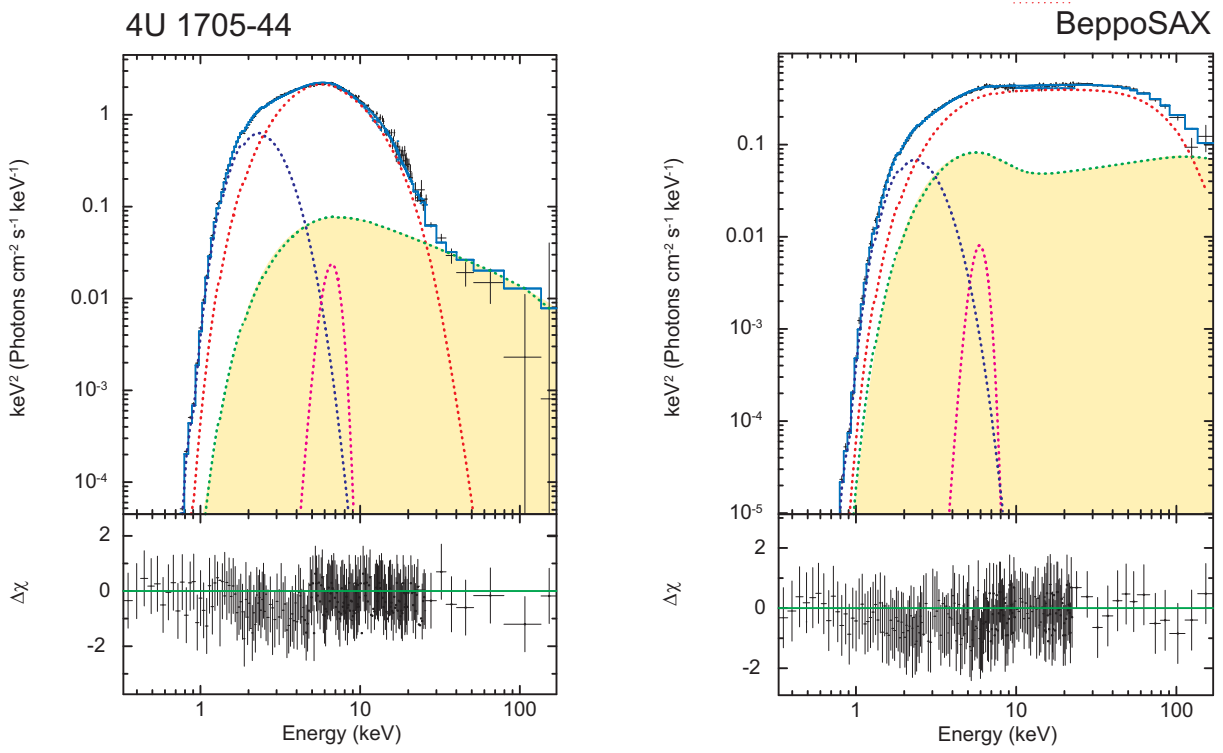}
\caption{Two representative $EF_E$ 
diagrams for {banana} and {island state} events 
of 4U~1705-44 using {\it BeppoSAX} data. 
Top:  best-fit spectra of 4U~1705-44 in $E*F(E)$ units using {\it BeppoSAX} observations  21292001 (left) 
 and
21292002 (right). 
The data  are presented by crosses, and the best-fit spectral  model   {\it wabs*(blackbody+Comptb1+Comptb2+Gaussian)} 
is presented by light-blue line. The model components  are shown by dark blue, red, green and crimson lines for 
{blackbody}, {Comptb1}, {Comptb2}  and  Gaussian components, respectively. 
{Bottom panels}: $\Delta \chi$ vs photon energy in keV. The best-fit model parameters for Banana state (left panel) 
are $\Gamma_1$=1.99$\pm$0.02, $kT^{(1)}_e$=2.47$\pm$0.01 keV, $\Gamma_2$=2.00$\pm$0.01, $kT^{(2)}_e$=46.0$\pm$0.7 keV 
and $E_{line}$=6.95$\pm$0.04 keV (reduced $\chi^2$=1.07 for 295 dof),  
while the best-fit model parameters for the island state (right panel) are 
$\Gamma_1$=2.00$\pm$0.04, $kT^{(1)}_e$=18.9$\pm$0.2 keV, $\Gamma_2$=2.01$\pm$0.03, $kT^{(2)}_e$=51$\pm$1 keV  
and $E_{line}$=6.51$\pm$0.08 keV (reduced $\chi^2$=1.02 for 296 dof) 
(see more details in Table 3).
}
\label{BeppoSAX_spectra}
\end{figure}

Using {\it BeppoSAX} data we demonstrate 
two 
$EF_E$ spectral diagrams for  
{banana} 
and 
{island branch} 
events 
along with the  {atoll} track presented in  Fig.~\ref{BeppoSAX_spectra}.    
We  show  
the best-fit $EF_E$ spectral diagrams  of 4U~1705-44 
using {\it BeppoSAX} observations  212921001 
(left panel) 
and
21292002 
(right panel). 
We indicate the data 
by crosses and the best-fit spectral  model (see details above)
by a light-blue line. 
 Red, green,  dark-blue and crimson lines are  for  
{Comptb1}, {Comptb2}, {Blackbody}  and {Gaussian} components, respectively. 
%
We present $\Delta \chi$ vs photon energy in keV in {the bottom panel}.
For the banana  
 (left panel)  
$\Gamma_1$=1.99$\pm$0.02, $kT^{(1)}_e$=2.47$\pm$0.01 keV, $\Gamma_2$=2.00$\pm$0.01, $kT^{(2)}_e$=46.0$\pm$0.7 keV, 
$kT_{BB}=0.50\pm 0.01$ keV  and $E_{line}$=6.50$\pm$0.03 keV (reduced $\chi_{red}^2$=1.12 for 132 dof),  
while 
for the island  (right panel)  
$\Gamma_1$=2.00$\pm$0.01, $kT^{(1)}_e$=18.9$\pm$0.1 keV, $\Gamma_2$=2.01$\pm$0.01, $kT^{(2)}_e$=51.0$\pm$0.9 keV, 
$kT_{BB}=0.54\pm 0.02$ keV and $E_{line}$=6.51$\pm$0.07 keV (reduced $\chi_{red}^2$=1.08 for 133 dof., see  details in Table 3). 
{The {BB} temperature $kT_{BB}$, independently of the  state, is around
0.6 ~keV (2 $\sigma$ upper limit).}   
The fit quality of the \sax\ spectra is significantly improved by inclusion  of this component. 
{
We aso find that in the soft state (ObsId=21292001)  $N_H$ is  a larger, namely $N_H=(2.43\pm0.03)\times10^{22}$ cm$^{-2}$ versus
 that in the hard state ($ObsId=21292002$) for which  $N_H=(2.16\pm0.04)\times10^{22}$ cm$^{-2}$.
}

Using our model we find that the photon  indices of  the {\it BeppoSAX} spectra 
$\Gamma_{1,2} = \alpha_{1,2}+1$ are 
 2.02$\pm$0.02 and 2.01$\pm$0.01 correspondingly (see Table 3).
We also find that the seed temperatures $kT_{s1}$ and $kT_{s2}$ vary  in the ranges 1.4$-$1.5 keV and 1.1$-$1.3 keV,  respectively.   

\subsubsection{{\it RXTE} data analysis}

The {\it RXTE} 
spectra  below 3 keV are not well calibrated;  
however one can use   the BB component parameters using  the {\it BeppoSAX} broad energy band (0.1-200 keV).
Thus,  in order to fit the {\it RXTE} data we have fixed   the {\it BB} temperature of the  
(inner disk) component at $kT_{BB}=$0.6 keV found  from our   analysis of   the \sax\ data .
{
Furthermore, because of the limited PCA low-energy coverage, we let 
$N_H$  be free and constrained  to lie  in the range $2 - 4\times 10^{22}$ cm$^{-2}$ as early reported by  different authors
(see e. g. Di Salvo et al. 2005, 2009; D'Ai et al. 20012; Egron et al. 2013).
}
In Table 4  we present the  best-fit spectral parameters  using the two-Comptb model applied to the {\it RXTE}  observations. 

\begin{figure}[ptbptbptb]
\includegraphics[scale=0.90,angle=0]{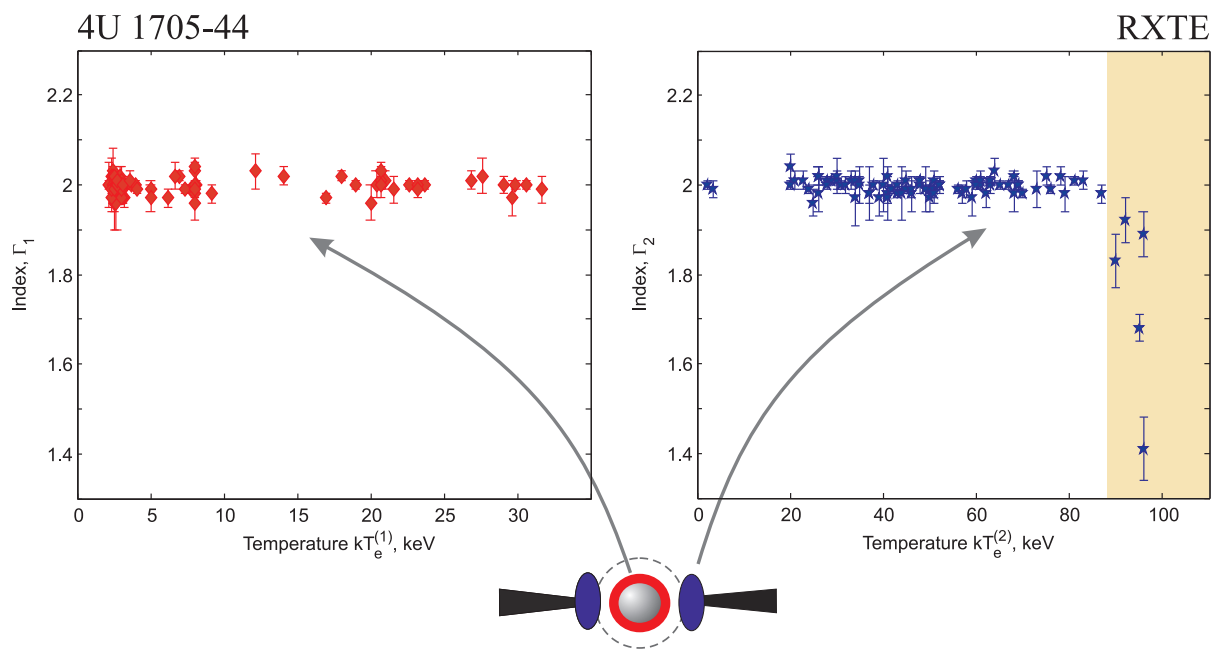}
\caption{
Photon indices $\Gamma_1$ and $\Gamma_2$ plotted 
vs. the best-fit electron 
temperatures of the Comptb1 and Comptb2 components, respectively,  measured in keV  
(see Table 4). 
Red and blue points 
correspond to Comptb1 and Comptb2 components, which are related to 
thermal upscattering of soft  photons by plasma electrons in  the NS boundary layer and CC, respectively. 
The range of the high electron temperatures $kT^{(2)}_{e}$ where $\Gamma_2<2$ is shaded yellow.
}
\label{index_temperature_12}
\end{figure}

 We obtain  that $N_{Com1}$
varies from 0.01 to 6.1 in units of $L_{37}/{D^2}_{10}$ ($L_{37}$ is the seed photon  blackbody luminosity 
in units  $10^{37}$ erg s$^{-1}$ while $D_{10}$ is distance in units of 10 kpc).  The   index $\Gamma_1$  is almost constant ($\Gamma_1=1.99\pm 0.06$) 
 for all observations (see Fig.~\ref{index_temperature_12}). 
 We reveal 
a two-phase behavoir  for the photon index $\Gamma_2$: the phase of the quasi-constant photon index ($\Gamma_2 = 2.01\pm 0.07$) when 
$kT^{(2)}_e$ varies between 3 keV and 85 keV, and the phase of the reduced photon index ($\Gamma_2<2$) for the high electron 
temperature $kT^{(2)}_e>85$ keV (see Fig.~\ref{index_temperature_12}, right diagram).  Note 
that the $Gaussian$ component width 
$\sigma_{line}$  does not significantly  change  and numerous tests show that it lies 
in the range 
of 0.5$-$0.8 keV.
\cite{D'Ai10}, \cite{diSalvo09}, \cite{Fiocchi07} and \cite{egron13} carried out   a detailed analysis of the XMM-$Newton$ spectra of 4U~1705-44  and they concluded  that the iron line is quite broad during all spectral states.
Therefore, we chose to fix $\sigma_{line}$ 
at a value 0.7 keV for all our fits. 
The electron temperatures $kT^{(1)}_{e}$ and $kT^{(2)}_{e}$ 
change in   wide ranges from 3 to  20 keV and from 3 to 100 keV 
respectively (see Figure~\ref{index_temperature_12}).

\subsubsection{\it Hardness$-$Flux Diagrams of 4U~1705-44 \label{ccd}}

The hardness$-$flux behavior depends on the energy bands.
Figure~\ref{3HID}  demonstrates the different shapes 
of the atoll tracks of 4U~1705-44 depending on the energy band (in flux units). 
In all these diagrams 
we use  the same observations but divided into  
different 
energy bands.  
Specifically, 
we plot the flux ratio 
[10-50 keV/3-10 keV] versus flux in [3-10 keV] (left), 
[10-50 keV] (center) and [50-200 keV] (right) 
 ranges 
measured in units of $10^{-9}$ erg s$^{-1}$cm$^{-2}$.  To make these diagrams,  we  apply our double Comptb spectral model.    

We found that the source exhibited  two color  states, namely, 
in the  IS and the BS. 
Here, the black arrow (left panel) indicates the direction of state changing from the island to the banana, whereas green arrow (center panel) marks the direction of state changing from the banana  to the island. 
 Note that 
{
this hysteresis 
}
effect has   
{
already  been pointed out  
}
 by BO02  for ID=20074 and 40051 sets. 
Previously, Gierl\`inski  \& Done (2002) also revealed  a  {\it Z-}shape of CCD track for 4U~1705-44. 

While the {atoll}-track for 4U~1705-44 
resembles a ``C''-track,  the source really moves in a more complicated manner 
resembling 
a {\it Z-}track. In fact, 4U~1705-44 is characterized  by an {extreme IS} , particularly well seen  as an upper alongated horizontal branch in the {\it central} panel. 
However this branch is not seen on the left panel (for a softer flux case).  
Thus, 
the source behaves as a {\it Z}-source
{
in terms of {hardness-flux} track. 
}

Note that  the hard emission events are  
 accompanied  by  a decrease of  radio emission  as observed  in 
{
a number of 
}
{\it Z-}sources [\cite{Hjellming90}].
However,  
 4U~1705-44 is not seen 
at radio wavelengths so far (see Fender \& Hendry 2000).
{
Furthermore, in agreement with previous timing analysis, 4U~1705-44 demonstrates the typical timing 
properties of atoll sources. 
But we should point out the unusual CCD/HID track of this standard atoll source, which is reminiscent of a   {\it Z}-shape in its { hardness-flux} diagram. This effect is essentially 
caused by its elongated extreme IS  
which is rarely observed   from  atoll sources.
}

\subsection{Overall pattern of X-ray properties \label{overall evolution}}
                                                                       	
\begin{figure}[ptbptbptb]
\includegraphics[scale=0.95, angle=0]{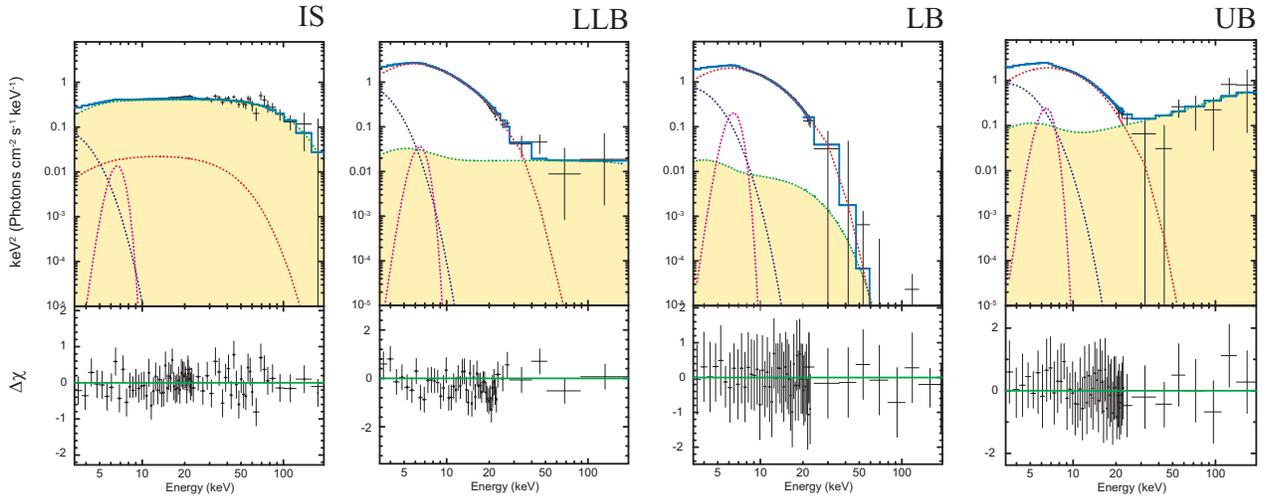}
\caption{Four representative $EF_E$  diagrams for different states along atoll-track of 4U~1705-44. 
Data are taken from {\it RXTE} observations 
40034-01-09-00 (island state, IS), 
40034-01-01-00 ({lower left banana}, LLB), 
40034-01-02-09 ({lower banana}, LB), 
and 40034-01-02-06 ({upper banana}, UB). 
The data are shown by black crosses, and the spectral model components are displayed  by dashed red, green, blue and purple lines for Comptb1, Comptb2, 
 Blackbody and Gaussian, respectively. Yellow shaded areas reveal an evolution of Comptb2 component during the spectral transition   (see also  Figs.~\ref{CCD} and \ref{index_temperature_12}).
}
\label{Zsp_compar_RXTE}
\end{figure}

\begin{figure}[ptbptbptb]
\includegraphics[scale=1.0,angle=0]{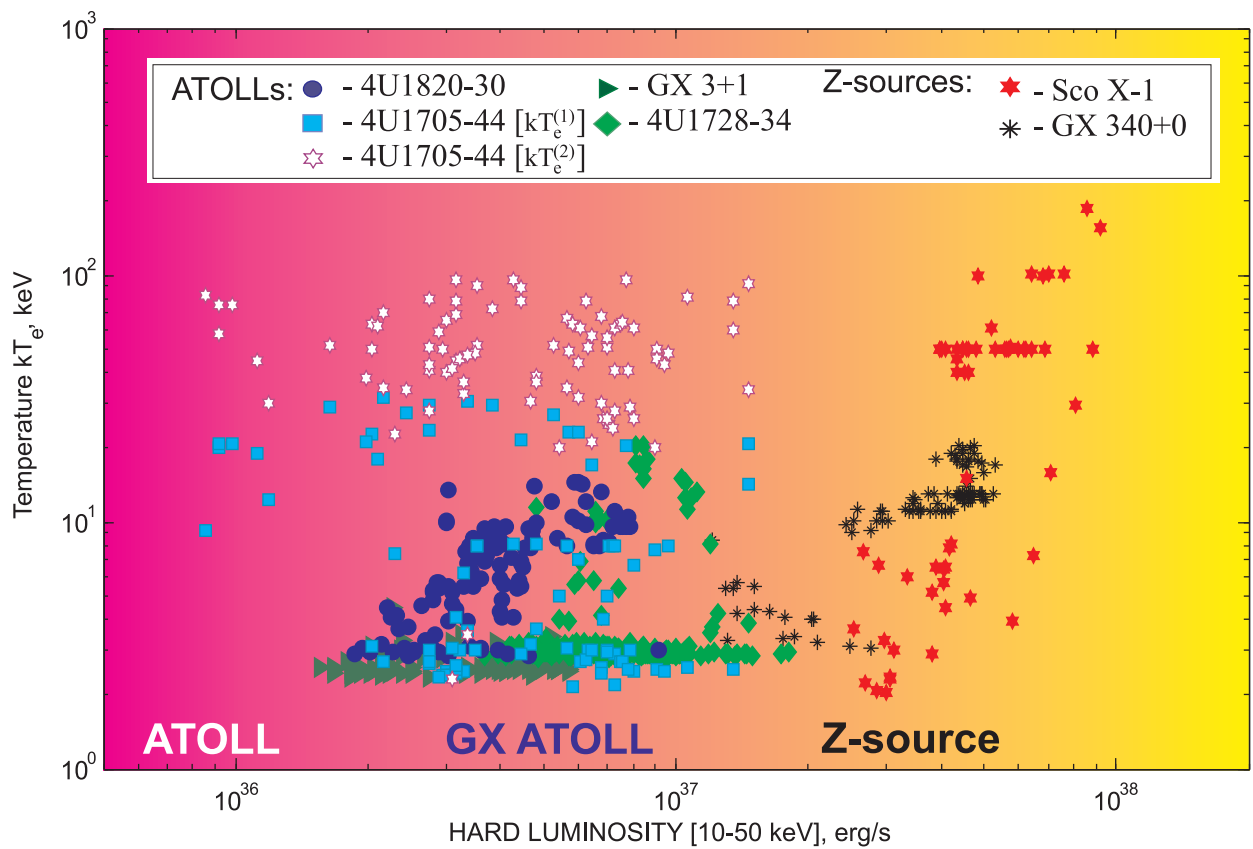}
\caption{
Electron temperature $kT_e$ vs luminosity in the 10-50 keV range, using {\it RXTE}  data for the 
{\it Z-}sources Sco~X-1 (red; TSS14), GX~340+0 (black; STF13) and  {atolls} 4U~1728-34 (bright~green;  ST11), GX~3+1 
(dark~green; ST12),  4U~1820-30 (dark~blue; TSF13) and 
4U~1705-44 [{\it light~blue} ($kT^{(1)}_e$), {\it white} ($kT^{(2)}_e$)]. 
}
\label{T_e vs lum_5obj}
\end{figure}

We have found that 4U~1705-44  was in four different spectral states, each of which 
is represented by $EF_E$ spectral 
diagrams are shown in Figure~\ref{Zsp_compar_RXTE}. 
Data (shown by black crosses) are taken from {\it RXTE} observations 
40034-01-09-00 (IS), 
40034-01-01-00 ({lower left banana}, LLB), 
40034-01-02-09 ({ lower banana}, LB), 
and 40034-01-02-06 ({\it upper banana}, UB). 
We show the spectral model components   by dashed red, green, 
blue and purple lines for Comptb1, Comptb2, BB  and {Gaussian}, respectively.     
 To  demonstrate an evolution of
Comptb2   between the {IS},
{LLB}, LB  
and {UB} 
branches, we use yellow shaded areas. 
Here we show  
 spectral changes 
in the energies greater than 30 keV that reflect   the 
contribution of the Comptonized components ({Comptb1} and {Comptb2})  for different   atoll branches.
In Figure~\ref{T_e vs lum_5obj} we also illustrate  these spectral changes using the plot of the electron temperature $kT_e$ versus the hard luminosity (for 10-50 keV energy range) 
along the total {atoll} cycle of 4U~1705-44. 

The  relative softening of the Comptb2 component  shown in the {LB} panel
in comparison with that presented in the {IS, LLB} and UB panels is seen in Fig. \ref{Zsp_compar_RXTE}.   On the other hand,  the {UB} panel shows the hardening of the spectrum at high energies, the photon index $\Gamma_{2}<2$ in this case.
 The hard emission at 50$-$150 keV (high-energy tail) becomes  stronger 
along the  UB, in comparison  of that for  the IS and LLB  (when $\Gamma_{2}\sim2$).
In our data set we have found five 
observations (20161-01-02-00, 20161-01-02-01, 40034-01-02-06, 40034-01-05-07, 40051-03-03-00) 
in which the {high energy tail}  
is strong and extended to 200 keV 
(similar to that presented in the right panel of Fig. \ref{Zsp_compar_RXTE}). 
 In all these cases the photon index of the $hard$ Comptonized component 
corresponds to  reduced values $\Gamma_2<2$. 
In Fig.~\ref{index_temperature_12} we combine these points with $\Gamma_2<2$ for $kT_e>80$ keV  with the nearly constant plateau of $\Gamma_2\sim2$ for $kT_e<80$ keV.

\begin{figure}[ptbptbptb]
\includegraphics[scale=1.0,angle=0]{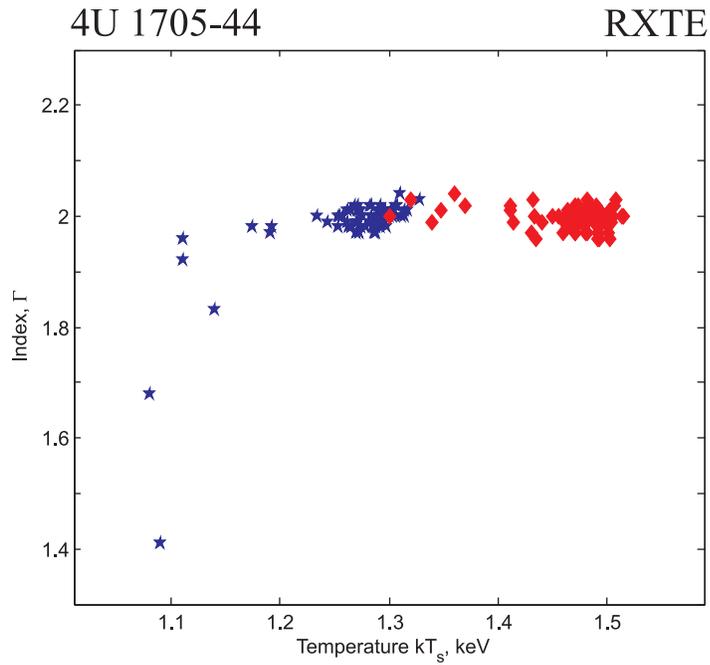}
\caption{
Photon indices $\Gamma_{1}$ and $\Gamma_{2}$ plotted vs.
the seed photon temperatures of the Comptb1 and Comptb2 components, respectively, measured in keV. 
(see more details in Table 4). 
Red and blue points correspond to Comptb1 and Comptb2 components which are related to the Comptonization of the soft  photons of the NS and disk  by the electrons  of the Compton cloud, respectively.
}
\label{index_temperature_s_12}
\end{figure}

To understand what are the possible reasons that can cause the above$-$mentioned hardening of the 4U~1705-44 spectrum at UB, 
we plot the photon indices $\Gamma_{1}$ and $\Gamma_{2}$  vs.
the seed photon temperatures $T_{s1}$ and $T_{s1}$ (see Fig. \ref{index_temperature_s_12}).  
Red and blue points correspond  to Comptb1 and Comptb2, respectively and thus one can clearly distinguish   the Comptonization of  soft  photons of the NS surface and that related  to   disk photons in the TL. 
As is seen from this plot,  $\Gamma_{2}$ is less than 2 when the $kT_{s2}$ monotonically drops from 1.3 keV to 1.1 keV. This is the first indication that the index drops when the  area 
of the TL increases. 

On the other hand, in Figure~\ref{index_temperature_12} we show the photon indices $\Gamma_1$ and $\Gamma_2$ 
versus $kT_e$ 
of the Comptb1 (red points) and Comptb2 (blue ponts) components
(see also Table 4).  
The {decreasing index phase} of the Comptonization  component is strongly related to the high electron temperature  ($kT^{{(2)}}_e>85$ keV).  That index remains constant
when the electron temperature $kT^{{(2)}}_e$ varies within the range  3$-$85 keV.

Our Comptonization model demonstrates a high quality  
for all data sets. 
The value of 
$\chi^2_{red}=\chi^2/N_{dof}$, where $N_{dof}$ is a number of degree of freedom, 
is   about 1.0. 
Only for
less than 2\% of the spectra does
$\chi^2_{red}$ reach 1.4. But $\chi^2_{red}$   never exceeds a  a rejection limit  of 1.5 which corresponds to  a 90\% confidence level. 



\subsubsection{Light curve and related spectral characteristics }

A complex behavior in a wide range of timescales, from seconds to years, is seen in the X-ray light curve of 4U~1705-44  (e.g., BO02). 
In this section  we present  the source variability
from hours to days. 
In Figure~\ref{lc_2000} we show the source and model parameter evolution for the whole set of the observed spectra.
Here, 
from the top to the  bottom, we present 
the evolution of the count rate (in counts s$^{-1}$) in the 2$-$9 keV range  using the 16~s time resolution;  of the {hardness ratio} HC [10$-$50 keV]/[3$-$10 keV];  the 
model flux in the 3$-$10 keV  and 10$-$50 keV energy ranges ({black and green} points, respectively);
$kT^{(1)}_e$ (red) and $kT^{(2)}_e$ (blue), the normalizations $N_{Com1}$ (red) and  $N_{Com2}$ (blue)
and   $\alpha=\Gamma-1$ for $Comptb1$ (red points) and $Comptb2$ (blue points)
for $R1-R3$ sets corresponding to  the time period from MJD 50,380 to MJD 51,650. 

The 
light-curve intervals, corresponding to values of $\Gamma_2 < 2$ 
are shown by blue vertical strips. 
For these phases 
the electron temperatures $kT^{(2)}_e$ is extremely high, while the {ratio} 
[10$-$50 keV]/[3$-$10 keV] 
coefficient (HC) is  arbitrary. 
It should be also noted   that  {a hardness ratio} is very sensitive to the choice of energy bands, as shown above 
(see Fig.~\ref{3HID}). 

All $hard\to soft$ transitions 
of 4U~1705-44 (see third panel from the top)  are related to  a
significant increase of the 3$-$10 keV flux and thus to  a decrease of the HC coefficient. 
Furthermore, one can also see from  Figure \ref{lc_2000}
how the electron temperature $kT^{(2)}_e$ (blue points) drops from 95 keV  to 5 keV 
during the hard-to-soft transitions (island-banana cycle).  $kT^{(2)}_e$ reaches  its maximum 
at the UB branch (e.g. at MJD 51,301). 
In  Fig.~\ref{Zsp_compar_RXTE}  we present the corresponding X-ray spectrum for this branch. 
Note that  
$kT^{(1)}_e$ (red points) varies from 3  to 20 keV, indicating that  the Comptonization component related to the innermost region of the transition layer, next to the NS surface, is much cooler than  the TL outer part related to $kT^{(2)}_e$. 

The  normalizations $N_{Com1}$ (red points) and $N_{Com2}$ (blue points) are only weakly  correlated
with  
a coefficient HC  ({hardness ratio color})   and  
 count rate in the 2-9 keV  band.  In fact,  for all spectral states 
the NS  emission always  dominates (see Figure~\ref{lc_2000}).  
$N_{Com2}$ is  variable (see Table 4).
Furthermore, $N_{Com2}$ (or $\dot M$) correlates with an increase of 
$T_e^{(2)}$  
   during X-ray flares (see  panels 2-4 from the bottom of Fig.~\ref{lc_2000}).

The  fractional contribution of  the Comptonized component associated with the seed disk photons 
(related to $N_{Com2}$)
is always much weaker  than that  contributed  by the  NS seed photons 
 (related to $N_{Com1}$).  In fact,  $N_{Com2}\sim 0.1N_{Com1}$).
As we show in the bottom panel  of 
 Figure \ref{lc_2000}, $\alpha_1$ and $\alpha_{2}$ 
($\Gamma_{1,2}=\alpha_{1,2}+1)$) only  slightly vary with time around 1, 
except of  for five points where $\alpha_{2}$ drops   
from this level.

{
The seed photon  temperatures $kT_{s1}$ and  $kT_{s2}$ 
increase  from 1.3 to 1.5 keV (for the $Comptb1$) and from  1.1  to 1.3 keV (for the $Comptb2$), 
respectively, during banana (soft state) $\to$ island (hard  state) transition. In order to evaluate the changes of a size  of the seed photon region  during the state transtion,
we use the parameters obtained by the Comptb models (see Tables 3 and 4).

Furthermore, to evaluate the radius of the illuminated TL region by the seed
photons that  are finally Comptonized, 
we assume that the seed photon emission is  a BB and the  total Comptonized flux 
is $F_{Com} = \eta F_{seed}$ where  $\eta$ is an enchancement Comptonization factor (which is 
$(4/5)[\log (T_e/T_s)+1.5]$ for the case $\alpha=1$ ($\Gamma=\alpha+1=2$), see \cite{st80}).
But 
$F_{seed}$ is obtained 
as
\begin{equation}
F_{seed} = \sigma T_{s,ef}^4 (R_{seed}/D)^2, 
\label{Fseed}
\end{equation}
and so an apparent radius of the seed photon area is defined by 
\begin{equation}
R_{seed} = 3\times 10^4 D \frac{\sqrt{F_{com}/\eta}}{T_{s,ef}^2} ~~~km, 
\label{Fseed}
\end{equation}
where $D$ is the distance in kpc, $T_{s,ef}\sim T_{s1}$ is in keV,  $F_{seed}=F_{Com}/\eta$,  $F_{Com}$  is the Comptonized radiation and $F_{seed}$ and $F_{Com}$ are  in erg cm$^{-2}$ s$^{-1}$.
By considering the $F_{Com}$  obtained using the {\it RXTE} spectra 
we find that  $F_{Com}\sim 9.64\times 10^{-9}$ and $9.5\times 10^{-10}$ erg cm$^{-2}$ s$^{-1}$ in the BS 
and IS, respectively. We use  a distance $D$ of 7.4 kpc, to obtain
$R_{seed}\sim$ 5.6 km
in the BS 
 and $R_{seed}\sim$ 1.6 km
in the IS 
state, respectively.  We should emphasize that the above estimates of  $R_{seed}$ are related to the effective area of the seed photons which is $4\pi R_{seed}^2$ is in our case. This area is definitely less than the NS area $4\pi R_{NS}^2$ as it should be.

As for the BB component, in the BS 
and IS,  the BB temperature ,  $T_{BB}$ is  about  
0.6 keV.
In turn, the BB radius $R_{BB}$ is derived via $L_{BB} = 4\pi R^2_{BB}\sigma T^4_{BB}$, where $L_{BB}$ is
the BB luminosity and $\sigma$ is the Stefan$-$Boltzmann constant. 
Assuming a distance $D$ of 7.4 kpc [see \cite{Haberl_Titarchuk95}] the BB region
has an apparent radius $R_{BB}$, 
{which varies from 5.5 km (in the IS) to 120 km (in the BS). These values are
}
in agreement with the emission coming from 
the AD (see Fig. \ref{geometry}). 
Thus, the radii $R_{seed}$ and $R_{BB}$ have reasonable values that are compatible with the adopted model (see Sect.~\ref{model choise}).
}




\section{Discussion}

\subsection{Comparative analysis  of   spectral properties of  
{\it  Z-} and  {Atoll} sources} 

In this section  we investigate 
how the X-ray spectral characteristics  depend on $\dot M$ 
variation 
relying on X-ray observations of  4U~1705-44  during its spectral  evolution 
across the banana-island branches. 
In fact, we  discover a new type of   $\Gamma-$index behavior  in  4U~1705-44 
with respect  to other known accreting NS sources. But variation of the index  in  4U~1705-44
is quite similar to that  observed  in the prototypical  {\it Z-}source Sco~X-1. Thus,  it is worth   comparing 4U~1705-44 with other NS binaries  and  identifying 
differences and similarities  of 
these sources (see Table 5).
We  make  a comparison 
with six 
sources: {\it Z-}sources GX~340+0  (STF13),  Sco~X-1 (STS14),   and  {\it atolls}  4U~1705-44, 4U~1728-34 (ST11) , GX~3+1 (ST12) and 
4U~1820-30  (TSF13),  using  the same double {\it Comptb} spectral model.
We also 
confront  the spectral evolution  of 4U~1705-44 
with that in 
 XTE~J1701-462 (NS source) since these two objects   
demonstrate  spectral properties 
common to {\it Z-}sources and  atolls  depending on their X-ray luminosity. 

\subsubsection{Approximate Constancy of the Photon index versus the Electron Temperature} 

{\it Z-}sources GX~340+0 and  Sco~X-1, and $atolls$, 4U~1728-34,  4U~1820-30, GX~3+1,   show a quite similar behavior  
of $\Gamma$ versus  $T_e$ at relatively low electron temperatures (3 keV $<kT_e<$ 60 keV). 
The  spectral  index $\Gamma_2$ is almost constant at a value of approximately  
about 2. 
This index behavior   can presumably indicate that, for these five sources,
  the accretted material releases its  gravitational energy 
in  the whole transition layer (TL) (see  ST11,  FT11,  ST12, STF13,  TSF13,  and TSS14).  
 According to  the TSS14  model,  this energy release  is  much higher in the TL than the cooling flow of the soft disk photons.  

\begin{figure}[ptbptbptb]
\includegraphics[scale=1.1,angle=0]{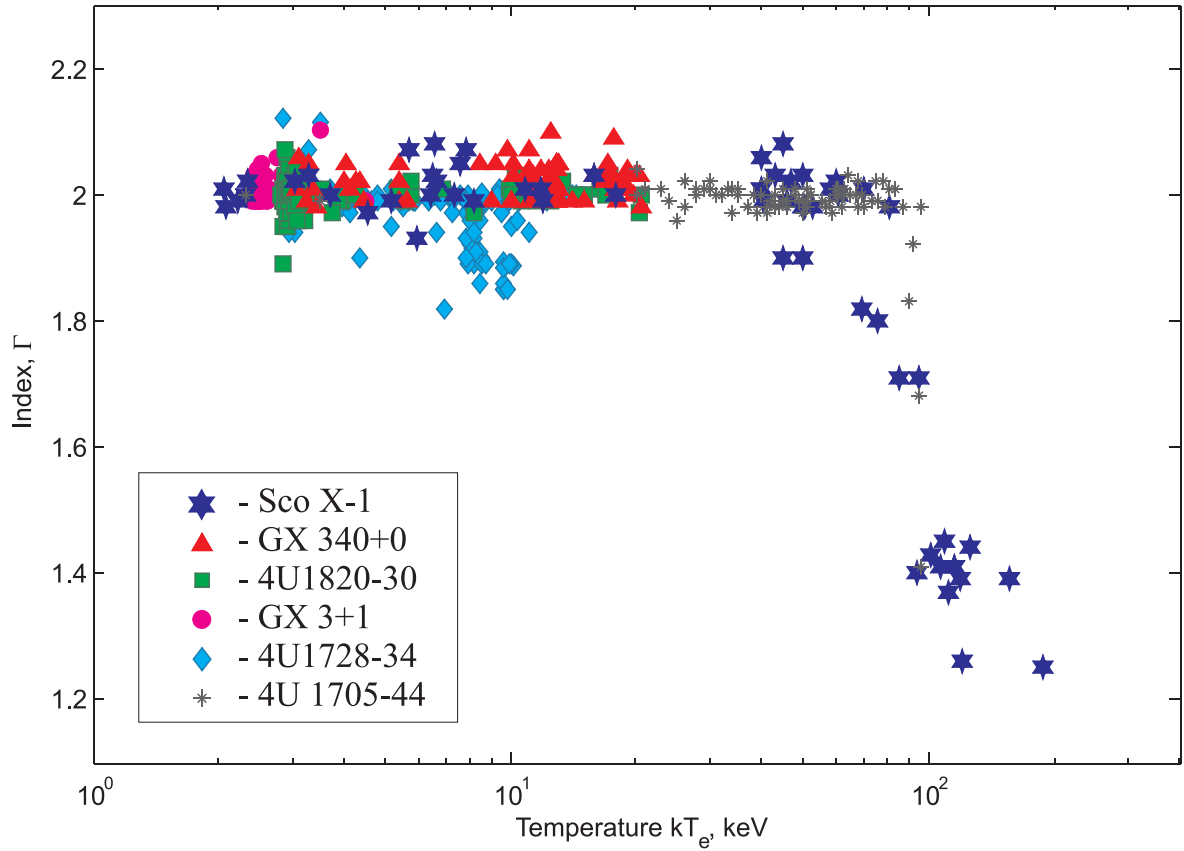}
\caption{
Photon 
index $\Gamma$ 
vs.
$kT_e$ 
for {\it Z}-sources Sco~X-1 (blue stars;  TSS14) and  GX~340+0 (red triangles;  STF13) and 
atoll sources 4U~1705-44 (black), 4U~1728-34 (bright~blue diamonds;  ST11), GX~3+1 (pink circles; ST12), and 4U~1820-30 (green squares, TSF13). 
}
\label{gam_te_5obj}
\end{figure}

For   Sco~X-1 (TSS14) and 4U~1705-44  
 we find   a   wider range of $T_e$
than that  in other {atolls}  and {\it Z-}sources.
Moreover, Sco~X-1 gives us a possibility 
to identify  the index sample  
when $L>L_{\rm Edd}$. 
It is worth noting that in this case the resulting luminosity is less than or equal to the critical one, whose value depends on the plasma (electron) temperature value.  We find 
that 
$\Gamma_2$ substantially   decreases when  
$kT_e>$ 85 keV (for 4U~1705-44 at UB) and $kT_e>$ 60 keV (for Sco~X-1 at FB) 
(see Fig.~\ref{gam_te_5obj}). 
In 4U~1705-44 the photon  index of the hard tail $\Gamma_2$ changes between 1.4 and 2 
in {upper banana} state. 
A similar case behavior  also takes place   in  Sco X-1   at FB
(see TSS14), 
where
the photon index $\Gamma_2$ varies between 1.3 and 2.
when the electron temperature 
of the TL  is high ($60$ keV $<kT_e<200$ keV). 

Note that 
{ 
only two sources, 4U~1705-44 and Sco X-1 demonstrate a decrease of $\Gamma_2$
at high $kT_e$
}
The decreasing index phase  of 4U~1705-44  begins at a higher electron temperature ($kT_e\sim 80$ keV) than that  in Sco~X-1 ($kT_e\sim 60$ keV). On the other hand, the upper limit of $T_e$ ranging is higher in Sco~X-1 ($kT^{max}_e$ exceeds 100 keV) 
than that for 4U~1705-44 ($kT^{max}_e\sim 95$ keV), which is possibly related to different X-ray luminosity levels (for  atolls and {\it Z-}sources). 
However, 
 4U~1705-44 emits  very close to a {\it Z-}source luminosity range as can be seen  from Fig.~\ref{T_e vs lum_5obj}. 

We contrast 
our results for {\it Z-}sources and  atolls with the spectral behavior of 
XTE~J1701-462 (see Lin et al. 2009,  hereafter LRH09). This source   evolves from the atoll to {\it Z-}stage  and finally   reaches the Eddington luminosity. 
It is surprising   that during the hard spectral state of XTE~J1701-462  the photon index  obtained by LHR09 is close  to  2,
while in the soft state
the index $\Gamma$ increases 
up to  2.5.  Note, that LRH09 investigate  the 
spectrum in the  3$-$80  keV energy range only.
 It is  possible that XTE~J1701-462 is a unique NS object, which  somehow differs from the standard atoll and {\it Z-}sources.
In fact, LHR09 use  various  models to  fit the data,
and their models take into account  
the role of a weak Comptonization 
in the soft state and 
in the hard state.

\section{CONCLUSIONS \label{summary}} 

We have investigated  the spectral 
property correlations with mass accretion rate found 
in 4U~1705-44 
using {\it BeppoSAX} and {\it RXTE}.
 We establish that the broadband energy spectra 
during all 
{atolls}  
can be  described  by  the double   {Comptb} additive model.
 These two Comptonized 
components are related to the different  {seed} photon temperatures ($T_{s1} \sim 1.5$ keV  and 
$T_{s2} =1.1-1.3$ keV).
To improve the fit quality we also include  an  iron-line ({Gaussian}) component. 
 
\begin{figure}[ptbptbptb]
\includegraphics[scale=1.0,angle=0]{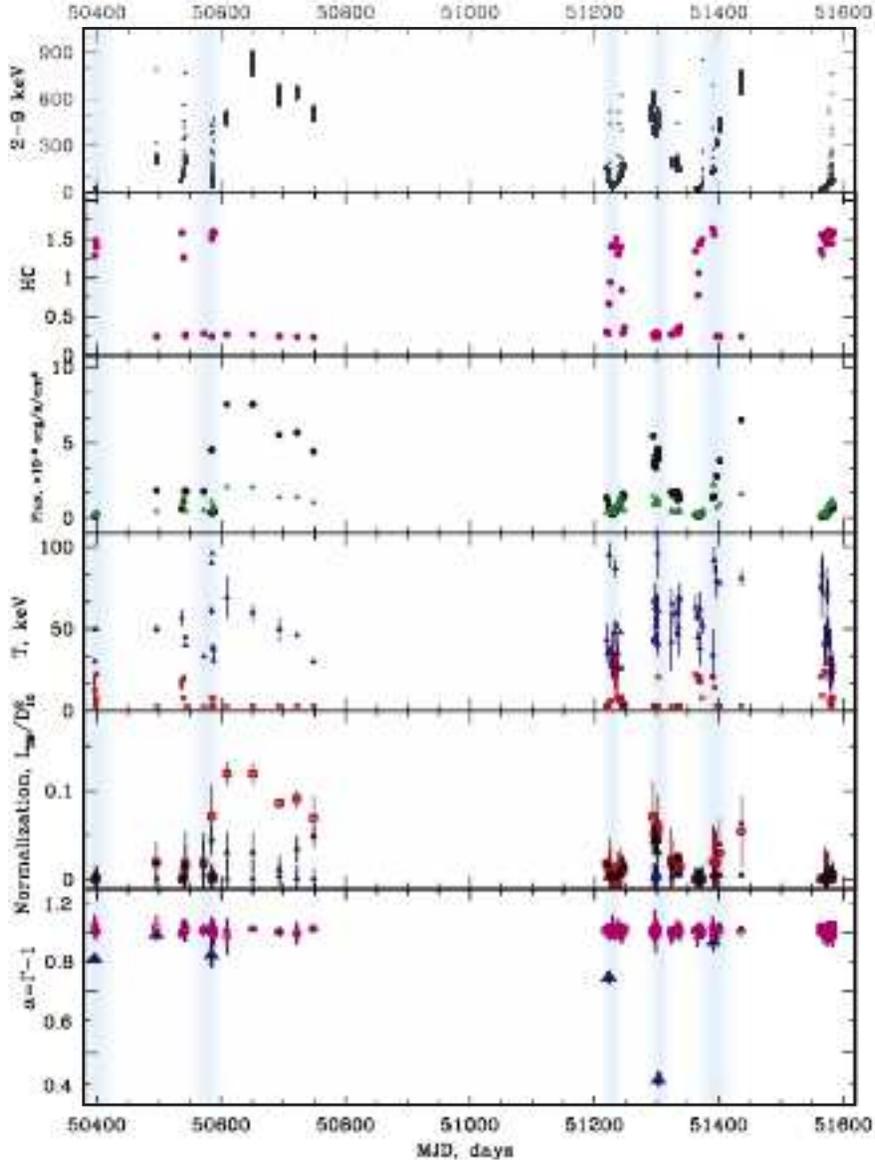}
\caption{{From Top to bottom:}
evolutions of  count rate [2-9 keV] in counts s$^{-1}$ with 16~s time resolution, 
the {hardness ratio} coefficient HC [10-50 keV]/[3-10 keV], 
the  model flux in the 3-10 keV and 10-50 keV energy ranges (black and green points, respectively),  
the electron temperatures $kT^{(1)}_e$ ($red$) and $kT^{(2)}_e$ ($blue$) 
in keV,  
blackbody normalizations of Comptb1 and Comptb2 ($red$ and blue, respectively),   
and  the spectral indices $\alpha_1$ and $\alpha_2$ 
(red and blue) for Comptb1 and Comptb2  components, respectively, for  1996$-$2000 evolution 
events (R1-R3 set). 
The 
phases of the light curve, related 
to the {reduced spectral index $\alpha_2$}, 
are marked with blue vertical strips. 
}
\label{lc_2000}
\end{figure}

Spectral  modeling 
4U~1705-44  gives us a possibility 
to separate 
two distinct   zones of the spectral formation along the {atoll} track. One is associated with   
the hard component while the other   is related to the soft component. 
We  find  that the hard-to-soft spectral transition is   driven by  mass accretion rate $\dot M$.  
In fact,  $N_{Com1}$ and  $N_{Com2}$  (normalization parameters; see Eq.~1) of the  Comptonized components  
are linerally related to mass accretion rate, $\dot M$.  
We establish  an increase of  $N_{Com2}$ (or $\dot M$),  
which correlates with an increase of the electron temperature $kT_e^{(2)}$ 
during X-ray flares  (see second and third  panels from the bottom of  Fig. \ref{lc_2000}).

We establish the stability of both photon indices $\Gamma_1$ and $\Gamma_2$ 
around 2 during the ISs and BSs, 
 which correspond to  the range of  the CC (TL)  electron temperature of   $kT^{(2)}_e$ 
 from 3 keV to  85 keV, while the decrease of 
$\Gamma_2$ is observed during the  high electron temperature phase ($kT^{(2)}_e>$ 85 keV). 
We explain the observed stability of $\Gamma_{1,2}$  
within
 2 
using  a model in which the
energy deposition takes place  
in the  TL and  the energy release   dominates the disk photon flux.
This result is similar to what was previously found in  the  atolls  
4U~1728-34, 4U~1820-30 and GX~3+1 and  {\it Z-}source GX~340+0 through all spectral states and in {\it Z-}source Sco~X-1  during HB$-$NB$-$botFB.  In addition to this  index plauteau phase,   of $\Gamma_{1,2}\sim2$   we also find the decreasing index $\Gamma_{2}$  {\it detected} over the UB 
(like that established  in   Sco~X-1 at FB).
 We interpret this index decrease  using the TSS14 model, where 
the  energy release occurs   only in some outer part of the TL.  
 In fact, the radiation pressure  halts 
the accretion flow 
in the innermost part of the TL, where the plasma temperature is effectively dictated 
by the NS surface photons.  Thus,    the electron (Klien-Nishina) cross-section   increases with diminishing  $T_e$
(see \S 4 of TSS14 for details). 

During this decreasing index stage, 
the  spectra of 4U~1705-44 and Sco~X-1 exhibit  an increase of $T_e$  
of the $hard$ Comptonization tail.
Note that  in BHs the index rises 
and then saturates when $\dot M$ increases. 
These   indices  behave  in a manner drastically different  from that seen in the NS  sources,  Sco X-1 and 4U~1705.  

We suggest that the robust nature of the parameter interdependencies resulting from the application of our model points to fundamental insights into the complex, and poorly understood, behavior or LMXBs.
  
{\it Acknowledgments:}
We recognize  a contribution of the referee who has thoroughly  checked the paper context.
N. Shaposhnikov also acknowledges the support of this research by the NASA 
under Grant  NNX13AF39G.


\newpage

%
%
\begin{deluxetable}{lcccc}
\tablewidth{0in}
\tabletypesize{\scriptsize}
    \tablecaption{The list of $BeppoSAX$ observations of 4U~1705-44  used in our analysis.}
    \renewcommand{\arraystretch}{1.2}
\tablehead{
Number of Set & Obs. ID& Start Time (UT)  & End Time (UT) &MJD Interval }
\startdata
S1 & 21292001  & 2000 Aug 20 16:20:16 & 2000 Aug 21 19:22:57 & 51776.6-51777.8 \\
S2 & 21292002  & 2000 Oct. 3 12:30:07  & 2000 Oct. 4 21:11:38  & 51820.5-51821.9 \\
      \enddata
   \label{tab:table}
References. 
(1) \citet{Fiocchi07};
(2) Piraino et al. (2007); 
(3) \citet{egron13}.

\end{deluxetable}


%
%
\newpage
\begin{deluxetable}{llll}
\tablewidth{0in}
\tabletypesize{\scriptsize}
    \tablecaption{The list of groups of {\it RXTE} observation of 4U~1705-44}
    \renewcommand{\arraystretch}{1.2}
\tablehead{Number of Set  & Dates, MJD & {\it RXTE} Proposal ID&  Dates UT \\
                          &            &                 &           }
 \startdata
R1  &    50,539-50,542 & 20073~(1)        & 1997~ Apr. 1$-$4 1997            \\
    &    50,495-50,748 & 20074~(1), (2)        & 1997~ Feb. 16$-$Oct. 27         \\
    &    50,398-50,587 & 20161~(1)        & 1996~Nov. 11$-$1997~May 19, 1997    \\
R2  &    51,290-51,583 & 40034~(1), (3) & 1999~Apr. 22$-$2000 Feb. 2     \\
R3  &    51,219-51,247 & 40051~(1), (2)     & 1999~Feb. 10$-$ Mar. 10         \\
      \hline
      \enddata
    \label{tab:par_bbody}
References:
(1) \cite{muno02};  
(2) \cite{barret02}; 
(3) Egron et al. (2013).
\end{deluxetable}

\begin{deluxetable}{llccccc}
\tablewidth{0in}
\tabletypesize{\scriptsize}
    \tablecaption{Best-fit Parameters of Comparative Spectral Analysis of {\it BeppoSAX} and {\it RXTE}
Observations of 4U~1705-44 in the 0.3$-$200~keV Energy Range Using Two 
Additive models$^{\dagger}$: 
{\tt wabs*(Blackbody + Comptb + powerlaw + Gaussian}) 
and  
{\tt wabs*(Blackbody + Comptb1 + Comptb2 + Gaussian}). 
}
    \renewcommand{\arraystretch}{1.2}
\tablehead{
       & Source state    &     IS   &        IS       &    LB    &       LB       &       UB        \\
  & Satellite    &$BeppoSAX$&       {\it RXTE}      &$BeppoSAX$&      {\it RXTE}      &      {\it RXTE}      \\
      \hline
Model & Parameter & 21292002 & 40034-01-07-02G & 21292001 & 40034-01-02-09 & 40034-01-02-06 }
 \startdata
wabs     & N$_H$ ($\times 10^{22}$ cm$^{-2}$) & 2.36$\pm$0.07  & 4.6$\pm$0.1 & 1.81$\pm$0.02& 4.65$\pm$0.02 & 4.82$\pm$0.07   \\
bbody    & kT$_{BB}$ (keV)   & 0.50$\pm$0.03  & 0.5$\pm$0.1 & 0.62$\pm$0.03 & 0.61$\pm$0.03  & 0.59$\pm$0.01  \\
     & N$_{BB}^{\dagger\dagger}$ & 3.8$\pm$0.3 & 7.8$\pm$0.3 & 2.7$\pm$0.4 & 100.7$\pm$0.9 & 14.9$\pm$0.2\\
Comptb     & $\alpha=\Gamma-1$& 1.00$\pm$0.01& 1.06$\pm$0.04 & 0.99$\pm$0.03 & 1.02$\pm$0.04 & 1.05$\pm$0.06 \\
     & kT$_{s}$ (keV)   & 1.31$\pm$0.08& 1.30$\pm$0.02  & 1.36$\pm$0.06& 1.28$\pm$0.09 & 1.30$\pm$0.07   \\
     & $\log(A)$           & 0.92$\pm$0.06& 0.95$\pm$0.02& -0.03$\pm$0.01&-0.24$\pm$0.09 & -0.22$\pm$0.02  \\
     & kT$_{e}$ (keV)   & 20.66$\pm$0.08 & 21.51$\pm$0.09 & 2.78$\pm$0.08& 2.86$\pm$0.09 & 2.78$\pm$0.03 \\
     & N$_{Com}^{\dagger\dagger}$ & 1.08$\pm$0.03 & 0.13$\pm$0.02&5.56$\pm$0.02& 6.62$\pm$0.01 & 7.16$\pm$0.07 \\
Power law & $\Gamma_{pow}$ & 1.88$\pm$0.05 & 1.5$\pm$0.4& 2.54$\pm$0.05 & 3.97$\pm$0.05 & 1.90$\pm$0.07 \\
     & N$_{pow}^{\dagger\dagger}$ & 2.95$\pm$0.02 & 2.13$\pm$0.01&1.26$\pm$0.02& 1.42$\pm$0.01 & 1.16$\pm$0.07 \\
Gaussian     & E$_{line}$ (keV)  & 6.45$\pm$0.08 & 6.48$\pm$0.09& 6.80$\pm$0.07 & 6.81$\pm$0.05 & 6.90$\pm$0.07 \\
     & $\sigma_{line}$ (keV)& 0.35$\pm$0.04 & 0.40$\pm$0.06& 0.39$\pm$0.07  & 0.41$\pm$0.03 & 0.41$\pm$0.05\\
     & N$_{line}^{\dagger\dagger}$ & 1.08$\pm$0.03 & 1.13$\pm$0.02&0.76$\pm$0.02& 0.72$\pm$0.01 & 0.66$\pm$0.07 \\
      \hline
     & $\chi_{red}^2$ (dof) & 3.28 (193)& 1.78 (79)  & 2.3 (200)& 1.34 (77) & 1.16 (77)  \\
      \hline
wabs     & N$_H$ ($\times 10^{22}$ cm$^{-2}$)  & 2.16$\pm$0.04  & 2.3$\pm$0.1  & 2.43$\pm$0.03 & 4.3$\pm$0.1  & 4.4$\pm$0.1   \\
bbody    & kT$_{BB}$ (keV)    & 0.54$\pm$0.07  & 0.60$\pm$0.04& 0.56$\pm$0.03 & 0.60$\pm$0.05& 0.60$\pm$0.04\\
     & N$_{BB}^{\dagger\dagger}$ &0.43$\pm$0.01 & 0.15$\pm$0.03  & 4.49$\pm$0.05 & 3.12$\pm$0.02& 3.16$\pm$0.05 \\
Comptb1&$\alpha_1=\Gamma_1-1$& 1.00$\pm$0.04  & 0.98$\pm$0.02& 0.99$\pm$0.07 & 1.01$\pm$0.02& 1.00$\pm$0.03 \\
     & kT$_{s1}$ (keV)       & 1.42$\pm$0.03  & 1.5$\pm$0.1  & 1.75$\pm$0.02 & 1.50$\pm$0.03& 1.50$\pm$0.04 \\
     & $\log(A_1)$              & 0.93$\pm$0.06 &2.0$^{\dagger\dagger\dagger}$& -0.04$\pm$0.01 &-4.02$\pm$0.07& 0.4$\pm$0.2\\
     & kT$^{(1)}_{e}$ (keV)  & 18.9$\pm$0.02  & 9.22$\pm$0.09& 2.47$\pm$0.04 & 2.71$\pm$0.09& 20.4$\pm$0.3 \\
     & N$_{Com1}^{\dagger\dagger}$  &1.08$\pm$0.02&0.04$\pm$0.02& 5.28$\pm$0.06 & 5.54$\pm$0.08& 4.58$\pm$0.04\\
Comptb2&$\alpha_2=\Gamma_2-1$& 1.01$\pm$0.03  & 1.01$\pm$0.02& 1.00$\pm$0.03 & 1.01$\pm$0.01& 0.41$\pm$0.07\\
     & kT$_{s2}$ (keV)       & 1.32$\pm$0.1   & 1.27$\pm$0.03& 1.29$\pm$0.05 & 1.26$\pm$0.02& 1.09$\pm$0.03\\
     & $\log(A_2)$              & -0.03$\pm$0.01 & 0.5$\pm$0.2  & -1.04$\pm$ 0.02 &-0.22$\pm$0.02 & -0.25$\pm$0.03 \\
     & kT$^{(2)}_{e}$ (keV)  & 51$\pm$1       & 83$\pm$5     & 2.04$\pm$0.08& 64$\pm$5& 96$\pm$7\\
     & N$_{Com2}^{\dagger\dagger}$  & 0.20$\pm$0.07&0.11$\pm$0.08&0.24$\pm$0.05&10.12$\pm$0.06&0.48$\pm$0.01\\
Gaussian & E$_{line}$ (keV)  & 6.51$\pm$0.08  & 6.40$\pm$0.05& 6.95$\pm$0.04 & 6.57$\pm$0.03& 6.67$\pm$0.03\\
     & $\sigma_{line}$ (keV)& 0.60$\pm$0.04  & 0.56$\pm$0.07& 0.79$\pm$0.08 & 0.64$\pm$0.07& 0.69$\pm$0.07 \\
& N$_{line}^{\dagger\dagger}$ & 1.07$\pm$0.09  & 1.09$\pm$0.08& 0.68$\pm$0.09 & 0.95$\pm$0.02& 1.15$\pm$0.04\\
      \hline
   & $\chi_{red}^2$ (dof) & 1.02 (296)     & 0.76 (86)   & 1.07 (295)  & 1.02 (86)   & 1.05 (86)  \\
      \hline
      \enddata
    \label{tab:BeppoSAX_fit_table}

$^\dagger$ Errors are given at the 90\% confidence level.
$^{\dagger\dagger}$ 
The normalization parameters of blackbody and Comptb components are in units of 
$L_{37}^{soft}/d^2_{10}$ 
$erg/s/kpc^2$, where $L_{37}^{soft}$ is the soft photon  luminosity in units of 10$^{37}$ erg s$^{-1}$, 
$d_{10}$ is the distance to the source in units of 10 kpc, 
the Gaussian component is in units of $10^{-3}\times total~~photons$ $cm^{-2}s^{-1}$ in line, 
$powerlaw$ component is in units of $10^{-2}\times ~~photons$ keV$^{-1}$ cm$^{-2}$ s$^{-1}$ at 1 keV;
$^{\dagger\dagger\dagger}$ when parameter $\log(A)\gg1$, it is fixed to a value 2.0 
for the model {\tt Comptb} 
(see comments in the text).              
%
\end{deluxetable}


%
%
\newpage
\bigskip
\begin{deluxetable}{lcccccccccccccc}
\rotate
\tablewidth{0in}
\tabletypesize{\scriptsize}
    \tablecaption{Best-fit Parameters of Spectral Analysis of PCA+HEXTE/{\it RXTE} 
Observations of 4U~1705-44 in the 3 -- 200~keV Energy Range$^{\dagger}$. 
Parameter errors are given at the 90\% confidence level.}
    \renewcommand{\arraystretch}{1.2}
 \tablehead                                                                                                          
{Observational & MJD, & $\alpha_1=$  & $kT^{(1)}_e,$ & $\log(A_1)$& $N_{Com1}^{\dagger\dagger\dagger}$ & $N_{Bbody}^{\dagger\dagger\dagger}$ & $kT_{s2}^{\dagger\dagger\dagger\dagger},$ & $\alpha_2=$  & $kT^{(2)}_e,$ & $\log(A_2)$& N$_{Com2}^{\dagger\dagger\dagger}$ &E$_{line}$,& $N_{line}^{\dagger\dagger\dagger}$ &  $\chi^2_{red}$ \\ 
ID             & (day)  & $\Gamma_1-1$ & (keV)           &            &                                    &                                      &  (keV)       & $\Gamma_2-1$ & (keV)           &            &                        & (keV)       &                                                &      (dof)}
 \startdata
20073-04-01-00 & 50539.56 & 0.99(1) & 7.72(8) & 2.00$^{\dagger\dagger}$ & 0.91(6) &  0.49(7) &  1.31(3) &  1.04(3) & 20(1)  & 1.7(1)   & 0.59(4) & 6.97(4) &0.28(8)& 1.45(87) \\ 
20073-04-01-01 & 50571.69 & 1.00(2) & 2.48(9) & 2.00$^{\dagger\dagger}$ & 1.92(3) &  1.5(1)  &  1.26(2) &  1.01(2) & 33(2)  & 1.77(4)  & 0.58(3) & 6.96(2) &0.24(9)& 0.91(87) \\ 
20073-04-01-02 & 50542.32 & 1.03(1) & 2.5(1)  & 2.00$^{\dagger\dagger}$ & 1.51(2) &  1.5(1)  &  1.23(4) &  1.00(6) & 40(1)  & -0.66(2) & 0.42(4) & 6.97(3) &0.25(4)& 0.79(87) \\ 
20073-04-01-03 & 50542.49 & 1.01(1) & 2.44(9) & 2.00$^{\dagger\dagger}$ & 1.86(1) &  1.6(2)  &  1.28(2) &  0.99(1) & 45(3)  & 0.43(5)  & 7.4(1)  & 6.94(4) &0.29(1)& 0.83(87) \\ 
20074-02-01-00 & 50495.75 & 1.02(4) & 2.38(7) & 2.00$^{\dagger\dagger}$ & 1.93(5) &  1.8(1)  &  1.19(3) &  0.97(3) & 50(1)  & 2.00$^{\dagger\dagger}$    & 0.01(1) & 6.95(7)  &0.31(2)& 0.87(86) \\ 
20074-02-02-00 & 50536.55 & 0.97(1) &16.9(2)  & 2.00$^{\dagger\dagger}$ & 0.08(2) &  3(1)    &  1.29(4) &  0.98(2) & 57(2)  & 2.00$^{\dagger\dagger}$    & 0.09(2) & 6.95(4) &5.1(9) & 0.75(88) \\ 
20074-02-03-00 & 50608.80 & 0.96(6) & 2.52(4) & 0.28(3)   & 11.9(2) &  3.0(1)  &  1.25(2) &  1.00(1) & 69(5)  & -0.63(2) & 8.4(7)  & 6.97(3) & 8(1)  & 0.92(88) \\ 
20074-02-04-00 & 50650.48 & 1.01(1) & 2.51(4) & 0.24(2)   & 11.9(4) &  3.05(2) &  1.26(6) &  1.01(1) & 60(2)  & -0.64(3) & 7.9(1)  & 6.95(4) & 8.2(3)& 0.91(88) \\ 
20074-02-05-00 & 50693.16 & 0.99(1) & 2.54(3) & 0.14(5)   & 8.6(3)  &  1.11(9) &  1.29(2) &  0.99(2) & 51(3)  & -0.61(4) & 0.39(1) & 6.74(5) & 6.4(2)& 0.81(88) \\ 
20074-02-06-00 & 50721.58 & 0.98(4) & 2.53(5) & 0.11(6)   & 9.1(2)  &  3.44(8) &  1.17(3) &  0.98(4) & 46(4)  & -0.13(2) & 0.13(3) & 6.65(3) & 7(1)  & 0.66(88) \\ 
20074-02-07-00 & 50748.19 & 1.01(1) & 2.59(6) & 0.07(4)   & 6.9(5)  &  4.90(7) &  1.31(4) &  1.01(2) & 30(1)  & 2.00$^{\dagger\dagger}$    & 0.1(1)  & 6.55(4)  & 6.2(4)& 0.76(88) \\ 
20161-01-01-002& 50398.35 & 0.99(3) & 3.12(6) & -2.3(8)   & 0.20(1) &  0.39(6) &  1.28(2) &  0.99(1) & 50(2)  & 0.43(2)  & 0.34(2) & 6.40(4) & 2.1(3)& 1.50(89) \\ 
20161-01-01-01 & 50396.35 & 1.03(4) & 12.2(4) & -1.9(6)   & 8(1)    &  0.25(4) &  1.26(3) &  1.02(4) & 30(1)  & 4.43(2)  & 0.13(1) & 6.41(3) & 1.2(1)& 0.96(92) \\ 
20161-01-01-03 & 50398.95 & 0.99(1) & 7.4(2)  & -1.8(5)   & 0.09(2) &  0.26(5) &  1.27(4) &  1.01(2) & 22.8(3)& 1.25(6)  & 0.30(2) & 6.42(2) & 2.6(2)& 1.27(89) \\ 
20161-01-02-00 & 50584.46 & 1.03(3) & 8.0(1)  &  2.00$^{\dagger\dagger}$ & 0.20(1) & 0.301(5)&  1.12(2) &  0.83(6) & 90(2)  & 0.66(6)  & 0.86(1) & 6.40(5) & 1.9(1)& 1.17(86) \\ 
20161-01-02-01 & 50585.46 & 1.00(1) & 8.1(2)  & -1.9(8)   & 0.10(3) &  0.41(2) &  1.27(3) &  0.98(5) & 95(8)  & 0.29(1)  & 0.59(3) & 6.41(9) & 2.8(3)& 1.26(90) \\ 
20161-01-02-02 & 50586.47 & 1.00(4) & 8.2(2)  & -1.8(6)   & 0.12(3) &  0.41(2) &  1.28(4) &  0.97(4) & 39(2)  & 1.29(1)  & 0.58(1) & 6.41(2) & 3.6(2)& 1.26(90) \\ 
20161-01-02-03 & 50587.35 & 0.97(2) & 3.2(1)  & -2.1(7)   & 0.08(1) &  0.37(1) &  1.31(3) &  1.00(1) & 31(2)  & 2.00$^{\dagger\dagger}$   & 0.63(2) & 6.39(5) & 2.9(2)& 1.25(89) \\ 
20161-01-02-04 & 50587.55 & 1.00(1) & 3.7(2)  & -1.8(5)   & 0.07(2) &  0.38(3) &  1.26(4) &  0.98(6) & 37(2)  & 2.00$^{\dagger\dagger}$   & 0.67(5) & 6.43(3) & 1.5(1)& 0.84(89) \\ 
40034-01-01-00 & 50584.46 & 0.99(2) & 2.9(3)  & -0.26(3)  & 7.12(7) &  4.49(9) &  1.29(2) &  1.01(2) & 61(4)  & -0.16(1) & 0.10(2) & 6.45(7) & 1.7(2)& 0.85(87) \\ 
40034-01-01-01 & 51294.33 & 1.00(2) & 2.50(6) &   0.45(1) & 7.10(8) &  4.43(7) &  1.19(3) &  0.98(1) & 43(1)  & -0.15(1) & 0.10(1) & 6.76(5) &0.11(1)& 1.15(86) \\ 
40034-01-01-02 & 51295.59 & 0.97(1) & 3.10(1) &  -0.17(1) & 5.19(1) &  4.49(9) &  1.27(4) &  1.00(1) & 67(2)  & -2.0(1)  & 0.20(5) & 6.52(5) &0.35(1)& 1.23(85) \\ 
40034-01-01-03 & 51297.53 & 1.00(5) & 2.14(2) &  -0.18(2) & 5.40(2) &  4.44(5) &  1.25(2) &  1.00(2) & 63(2)  & -1.95(3) & 0.10(1) & 6.61(5) &0.54(3)& 1.12(85) \\ 
40034-01-01-04 & 51297.47 & 1.02(6) & 2.43(1) &  -0.28(1) & 5.13(2) &  4.47(4) &  1.28(1) &  1.02(1) & 68(4)  & -0.16(2) & 0.60(1) & 6.40(4) &0.25(2)& 1.18(86) \\ 
40034-01-02-00 & 51298.15 & 1.01(1) & 3.04(3) &  -0.25(3) & 4.54(2) &  4.48(5) &  1.26(3) &  0.99(2) & 51(3)  & -0.14(1) & 0.65(3) & 6.59(5) &0.14(1)& 1.19(86) \\ 
40034-01-02-01 & 51298.69 & 0.96(6) & 2.69(1) &  -0.27(1) & 5.86(2) &  4.46(6) &  1.29(3) &  1.01(1) & 63(2)  & -0.18(2) & 0.63(1) & 6.60(3) &0.58(1)& 1.02(86) \\ 
40034-01-02-02 & 51298.76 & 1.00(1) & 3.04(1) &  -0.27(1) & 5.27(3) &  4.43(5) &  1.25(2) &  0.98(2) & 51(3)  & -0.13(1) & 0.67(1) & 6.44(5) &0.19(1)& 1.15(86) \\ 
40034-01-02-03 & 51298.83 & 0.99(3) & 2.72(2) &  -0.26(3) & 5.90(3) &  4.45(3) &  1.32(4) &  1.03(3) & 64(5)  & -0.16(2) & 0.62(1) & 6.71(3) &0.53(3)& 0.95(86) \\ 
40034-01-02-04 & 51298.89 & 1.02(2) & 2.97(1) &  -0.29(2) & 5.41(3) &  4.44(5) &  1.29(1) &  0.99(1) & 56(7)  & -0.15(2) & 0.61(3) & 6.48(5) &0.29(2)& 1.11(86) \\ 
40034-01-02-05 & 51301.53 & 1.00(1) & 2.18(4) &  -0.28(1) & 5.73(1) &  4.45(2) &  1.28(3) &  0.98(2) & 41(3)  & -0.16(2) & 0.65(1) & 6.55(8) &0.34(2)& 1.23(86) \\ 
40034-01-02-06 & 51301.69 & 1.00(3) & 20.4(3) &   0.4(2)  & 4.58(4) &  3.16(5) &  1.09(3) &  0.41(7) & 96(7)  & -0.25(3) & 0.48(1) & 6.67(3) &1.15(4)& 1.05(86) \\ 
40034-01-02-07 & 51301.69 & 1.00(1) & 2.48(5) &   0.23(6) & 6.13(1) &  3.04(6) &  1.29(2) &  1.00(1) & 61(3)  & -0.15(2) & 0.42(3) & 6.56(5) &1.19(5)& 1.00(86) \\ 
40034-01-02-08 & 51301.75 & 1.01(2) & 2.52(3) &   0.17(5) & 5.95(7) &  3.05(5) &  1.28(3) &  1.02(2) & 43(1)  & -0.18(1) & 0.41(1) & 6.66(7) &1.14(3)& 1.06(86) \\ 
40034-01-02-09 & 51301.83 & 1.01(2) & 2.71(9) &  -4.02(7) & 5.54(8) &  3.12(2) &  1.26(2) &  1.01(1) & 64(5)  & -0.22(2) & 0.12(6) & 6.57(3) &0.95(2)& 1.02(86) \\ 
40034-01-03-00 & 51324.06 & 0.97(3) & 2.38(3) &   0.17(7) & 2.24(7) &  1.44(6) &  1.28(4) &  0.99(1) & 42(7)  & -0.26(2) & 0.31(3) & 6.66(5) &0.40(1)& 1.00(86) \\ 
40034-01-03-01 & 51324.20 & 1.00(1) & 2.41(7) &   0.22(9) & 2.03(6) &  1.45(3) &  1.29(1) &  1.00(1) & 65(4)  & -0.22(1) & 0.34(1) & 6.40(5) &0.44(2)& 0.95(86) \\ 
40034-01-03-02 & 51325.17 & 0.99(2) & 2.34(5) &   0.43(8) & 1.88(5) &  1.46(5) &  1.27(3) &  0.97(4) & 59(3)  & -0.21(2) & 0.32(1) & 6.40(7) &0.42(1)& 0.97(86) \\ 
40034-01-04-00 & 51362.67 & 1.00(1) & 22.6(3) &   2.00$^{\dagger\dagger}$ & 0.08(2) & 0.45(2) &  1.30(2) &  1.01(1) & 63(4)  &  0.26(8) & 0.3041) & 6.45(3) & 2.8(6)& 1.14(86) \\ 
40034-01-04-01 & 51366.14 & 0.96(4) & 20.0(9) &   2.00$^{\dagger\dagger}$ & 0.04(2) & 0.04(2) &  1.28(3) &  0.99(2) & 58(1)  & -0.42(9) & 0.32(1) & 6.48(5) & 1.3(4)& 1.21(86) \\ 
40034-01-04-02 & 51367.14 & 1.00(1) & 19.0(8) &   2.00$^{\dagger\dagger}$ & 0.01(1) & 0.01(1) &  1.29(5) &  1.01(2) & 45(2)  &  -0.43(9)& 0.34(3) & 6.40(5) & 1.1(2)& 1.22(86) \\ 
40034-01-04-03 & 51369.20 & 1.01(2) & 21(1)   &   2.00$^{\dagger\dagger}$ & 0.05(2) & 0.02(1) &  1.30(4) &  1.00(1) & 38(4)  & -0.42(9) & 0.37(2) & 6.41(7) & 1.2(1)& 1.24(86) \\ 
40034-01-04-04 & 51370.20 & 1.02(1) & 18.0(9) &   2.00$^{\dagger\dagger}$ & 0.02(1) & 0.01(1) &  1.27(3) &  0.98(3) & 62(4)  & -0.4(1)  & 0.38(5) & 6.49(3) & 1.2(1)& 1.23(86) \\ 
40034-01-04-05 & 51373.46 & 0.98(2) & 8.0(5)  &   2.00$^{\dagger\dagger}$ & 0.03(1) & 0.01(1) &  1.31(4) &  1.00(2) & 52(3)  & -0.51(6) & 0.62(2) & 6.40(6) & 1.3(2)& 1.17(86) \\ 
40034-01-05-00 & 51333.24 & 1.01(3) & 3.0(1)  & -0.23(2)  & 2.51(7) &  2.40(1) &  1.27(3) &  1.01(1) & 48(1)  & -0.14(6) & 0.63(4) & 6.53(5) &0.16(2)& 1.14(86) \\ 
40034-01-05-01 & 51334.42 & 1.00(2) & 3.2(1)  & -0.22(1)  & 2.28(6) &  2.58(1) &  1.24(2) &  0.99(2) & 46(3)  & -0.14(6) & 0.61(5) & 6.60(5) &0.15(1)& 0.88(86) \\ 
40034-01-05-02 & 51335.23 & 0.97(1) & 3.0(4)  & -0.26(1)  & 1.89(2) &  2.30(1) &  1.29(3) &  1.01(2) & 51(7)  & -0.14(6) & 0.62(2) & 6.61(5) & 5.1(4)& 0.76(86) \\ 
40034-01-05-03 & 51336.09 & 1.02(2) & 2.51(6) & -0.08(2)  & 1.43(3) &  0.58(1) &  1.28(4) &  1.00(1) & 69(3)  & -0.37(6) & 0.85(3) & 6.50(3) & 4.9(7)& 0.79(86) \\ 
40034-01-05-04 & 51337.46 & 1.00(1) & 2.60(5) & -0.12(3)  & 1.43(1) &  0.57(1) &  1.26(2) &  0.98(4) & 68.0   & -0.36(6) & 0.83(1) & 6.53(6) & 3.8(4)& 0.84(86) \\ 
40034-01-05-05 & 51390.13 & 1.01(4) & 20.7(5) &   2.00$^{\dagger\dagger}$ & 2.00(8) & 0.35(4) &  1.27(3) &  0.97(6) & 34(6)  & -2.1(9)  & 0.45(3) & 6.40(5) & 4.7(4)& 1.07(86) \\ 
40034-01-05-06 & 51333.05 & 0.99(1) & 2.67(6) & -0.17(3)  & 1.44(1) &  0.58(1) &  1.25(4) &  1.00(1) & 60(4)  & -0.36(6) & 0.65(1) & 6.51(4) & 2.1(3)& 0.89(86) \\ 
40034-01-05-07 & 51392.27 & 1.02(2) & 14.1(9) &   2.00$^{\dagger\dagger}$ & 2.00(8) & 0.78(4) &  1.11(2) &  0.92(5) & 92(3)  & -1.10(9) & 0.37(3) & 6.56(3) &0.30(2)& 1.13(86) \\ 
40034-01-06-00 & 51396.23 & 1.00(1) & 2.91(5) &   2.00$^{\dagger\dagger}$ & 1.90(7) & 0.58(1) &  1.27(2) &  0.98(4) & 79(4)  & -2.00(6) & 3.17(1) & 6.67(5) & 4.3(4)& 1.08(86) \\ 
40034-01-06-01 & 51436.15 & 0.99(1) & 2.59(2) &   2.00$^{\dagger\dagger}$ & 5.42(8) & 0.54(5) &  1.29(3) &  1.01(1) & 81(2)  & -2.1(2)  & 6.32(1) & 6.95(7) & 9.0(4)& 1.08(86) \\ 
40034-01-06-09 & 51401.39 & 1.00(3) & 2.76(4) &   2.00$^{\dagger\dagger}$ & 2.95(8) & 0.56(3) &  1.30(1) &  1.02(2) & 78(5)  & -2.2(3)  & 3.99(8) & 6.68(6) & 5.2(4)& 1.18(86) \\ 
40034-01-07-00 & 51564.13 & 1.03(1) & 20.7(7) &   2.00$^{\dagger\dagger}$ & 0.06(1) & 0.18(1) &  1.28(2) &  0.99(1) & 76(3)  & 0.2(1)   & 0.11(3) & 6.43(5) & 1.3(5)& 1.06(86) \\ 
40034-01-07-01 & 51565.33 & 1.00(1) & 20.7(5) &   2.00$^{\dagger\dagger}$ & 0.08(1) & 0.16(1) &  1.29(2) &  1.02(2) & 75(7)  & 0.2(1)   & 7(1)    & 6.41(3) & 1.1(3)& 1.07(86) \\ 
40034-01-07-02G& 51566.26 & 0.98(2) & 9.22(9) &   2.00$^{\dagger\dagger}$ & 0.04(2) & 0.15(3) &  1.27(3) &  1.01(2) & 83(5)  & 0.5(2)   & 0.11(8) & 6.40(5) & 1.0(1)& 0.76(86) \\ 
40034-01-07-03 & 51573.72 & 0.99(3) & 31.7(5) &   2.00$^{\dagger\dagger}$ & 0.04(1) & 0.28(6) &  1.26(4) &  0.98(1) & 70(7)  & 0.71(7)  & 0.26(9) & 6.40(5) & 1.2(2)& 1.09(86) \\ 
40034-01-07-04 & 51569.32 & 1.00(1) & 23.7(8) &   2.00$^{\dagger\dagger}$ & 0.12(7) & 0.27(5) &  1.28(2) &  0.97(5) & 41(4)  & 0.57(9)  & 0.33(8) & 6.42(6) & 1.3(1)& 1.01(86) \\ 
40034-01-07-05 & 51574.65 & 0.97(4) & 29.7(9) &   2.00$^{\dagger\dagger}$ & 0.16(6) & 0.32(3) &  1.27(3) &  0.99(4) & 73(3)  & 0.69(7)  & 0.44(9) & 6.40(3) & 1.4(2)& 1.07(86) \\ 
40034-01-08-00 & 51572.65 & 1.00(2) & 29(1)   &   2.00$^{\dagger\dagger}$ & 0.07(6) & 0.23(5) &  1.29(2) &  1.00(1) & 52(7)  & 0.71(8)  & 0.20(9) & 6.43(5) & 1.1(1)& 1.09(86) \\ 
40034-01-08-01 & 51575.90 & 1.00(1) & 23.1(1) &   2.00$^{\dagger\dagger}$ & 0.23(5) & 0.43(6) &  1.27(4) &  0.99(5) & 49(3)  & 0.90(5)  & 0.63(9) & 6.41(5) & 1.8(5)& 0.79(86) \\ 
40034-01-08-02 & 51577.03 & 0.99(2) & 23.2(9) &   2.00$^{\dagger\dagger}$ & 0.24(6) & 0.45(5) &  1.26(3) &  0.98(6) & 44(2)  & 0.82(5)  & 0.65(7) & 6.44(6) & 3.2(1)& 1.15(86) \\ 
40034-01-09-00 & 51581.76 & 1.04(1) & 8.0(3)  &   2.00$^{\dagger\dagger}$ & 0.11(1) & 0.46(1) &  1.11(2) &  0.96(3) & 25(4)  & 1.06(7)  & 1.01(2) & 6.40(3) & 6.1(3)& 0.88(86) \\ 
40034-01-09-01 & 51580.84 & 1.00(1) & 3.0(3)  &   2.00$^{\dagger\dagger}$ & 0.21(4) & 0.47(3) &  1.30(3) &  1.01(2) & 21(5)  & 1.30(8)  & 0.89(1) & 6.41(5) & 3.2(2)& 0.85(86) \\ 
40034-01-09-02 & 51578.78 & 0.99(2) & 5.0(2)  &   2.00$^{\dagger\dagger}$ & 0.52(2) & 0.46(4) &  1.29(4) &  1.00(1) & 20(3)  & 1.14(9)  & 0.76(1) & 6.42(6) & 3.1(1)& 1.04(86) \\ 
40034-01-09-03 & 51583.10 & 1.00(3) & 8.0(4)  &   2.00$^{\dagger\dagger}$ & 0.10(1) & 0.52(3) &  1.27(1) &  0.99(1) & 24(2)  & 1.29(8)  & 0.99(3) & 6.40(2) & 6.3(3)& 1.16(86) \\ 
40034-01-09-04 & 51580.04 & 1.02(1) & 7.0(3)  &   2.00$^{\dagger\dagger}$ & 0.20(3) & 0.43(3) &  1.29(1) &  1.00(2) & 32(7)  & 1.09(9)  & 0.84(7) & 6.41(3) & 4.7(2)& 0.94(86) \\ 
40034-01-09-05 & 51581.68 & 0.99(3) & 3.0(4)  &   2.00$^{\dagger\dagger}$ & 0.10(1) & 0.37(4) &  1.31(2) &  1.01(1) & 29(3)  & 0.65(8)  & 1.27(2) & 6.40(2) &0.13(1)& 1.18(86) \\ 
40034-01-09-07 & 51581.69 & 1.00(1) & 4.0(2)  &   2.00$^{\dagger\dagger}$ & 0.45(1) & 0.46(3) &  1.30(1) &  1.02(2) & 26(4)  & 1.21(9)  & 0.96(1) & 6.41(2) & 5.4(2)& 1.18(86) \\ 
40034-01-09-08 & 51583.17 & 0.96(4) & 8.0(3)  &   2.00$^{\dagger\dagger}$ & 0.54(2) & 0.45(3) &  1.29(3) &  1.01(1) & 28(2)  & 1.16(8)  & 1.02(1) & 6.43(6) & 4.3(3)& 1.06(86) \\ 
40051-03-01-00 & 51219.60 & 1.00(1) & 2.7(1)  & 0.11(1)   & 1.82(1) &  1.92(1) &  1.30(2) &  1.00(1) & 43(4)  & -0.16(8) & 0.19(2) & 6.64(2) &0.11(1)& 0.99(86) \\ 
40051-03-02-00 & 51221.60 & 1.01(2) & 2.7(2)  &   2.00$^{\dagger\dagger}$ & 1.56(2) & 1.54(1) &  1.29(4) &  1.01(1) & 35(2)  & -0.13(8) & 0.18(2) & 6.51(6) &0.11(1)& 1.12(86) \\ 
40051-03-03-00 & 51223.53 & 0.99(1) & 4.11(9) &  0.41(8)  & 0.65(8) &  0.67(2) &  1.08(2) &  0.68(3) & 95(3)  & -0.14(8) & 0.46(9) & 6.40(3) &0.18(1)& 1.17(86) \\ 
40051-03-04-00 & 51225.46 & 0.97(2) & 6.2(1)  & 1.27(9)   & 0.40(6) &  0.38(2) &  1.29(2) &  0.98(2) & 37(1)  & -0.17(8) & 0.46(6) & 6.41(2) &0.12(1)& 1.03(86) \\ 
40051-03-05-00 & 51227.34 & 1.00(1) & 29.9(4) &   2.00$^{\dagger\dagger}$ & 0.30(4) &  0.34(2) &  1.28(1) &  1.00(1) & 28(3)  & -0.16(8) & 0.19(5) & 6.40(6) & 4.8(2)& 1.04(86) \\ 
40051-03-06-00 & 51229.59 & 1.02(4) & 27.6(2) &   2.00$^{\dagger\dagger}$ & 0.27(1) &  0.29(1) &  1.30(4) &  1.01(1) & 34(5)  & -0.12(8) & 0.15(4) & 6.45(2) & 2.3(3)& 0.95(86) \\ 
40051-03-07-00 & 51231.46 & 1.00(1) & 30(2)   &   2.00$^{\dagger\dagger}$ & 0.36(3) &  0.33(1) &  1.29(3) &  1.00(2) & 48(3)  & -0.16(8) & 0.24(3) & 6.40(2) & 5.6(4)& 0.93(86) \\ 
40051-03-08-00 & 51233.27 & 0.99(3) & 21(3)   &   2.00$^{\dagger\dagger}$ & 0.36(2) &  0.37(2) &  1.27(2) &  0.98(2) & 87(2)  & -0.09(4) & 0.31(3) & 6.47(3) & 6.1(3)& 1.06(86) \\ 
40051-03-09-00 & 51235.52 & 1.01(2) & 26(2)   &   2.00$^{\dagger\dagger}$ & 0.53(3) &  0.41(5) &  1.28(4) &  1.00(1) & 52(1)  & 2.00$^{\dagger\dagger}$      & 0.35(5) & 6.40(2) & 6.2(3)& 1.11(86) \\ 
40051-03-10-00 & 51237.52 & 1.00(1) & 8.0(3)  &   2.00$^{\dagger\dagger}$ & 0.60(5) &  0.50(2) &  1.29(2) &  1.00(4) & 35(3)  & 0.21(3)  & 0.46(4) & 6.42(6) & 9.5(4)& 0.89(86) \\ 
40051-03-11-00 & 51239.40 & 1.02(3) & 6.70(4) &   2.00$^{\dagger\dagger}$ & 0.71(4) &  0.53(3) &  1.27(3) &  1.02(2) & 26(1)  & 0.34(3)  & 0.78(6) & 6.48(2) &0.16(1)& 0.97(86) \\ 
40051-03-12-00 & 51241.86 & 0.99(1) & 8.1(2)  &   2.00$^{\dagger\dagger}$ & 0.58(6) &  0.58(2) &  1.29(1) &  1.01(1) & 48(2)  & 0.54(6)  & 1.07(4) & 6.40(3) &0.22(1)& 1.02(86) \\ 
40051-03-13-00 & 51243.93 & 0.97(3) & 5.0(3)  & -0.67(3)  & 0.78(4) &  0.83(2) &  1.28(2) &  0.98(4) & 26(1)  & 0.96(4)  & 1.13(3) & 6.59(2) &0.32(3)& 1.07(86) \\ 
40051-03-14-00 & 51245.93 & 1.00(1) & 3.10(7) &  0.90(8)  & 1.13(2) &  2.03(4) &  1.30(1) &  1.00(1) & 2.32(7)& -0.20(3) & 1.20(7) & 6.61(2) &0.27(2)& 1.11(86) \\ 
40051-03-15-00 & 51247.85 & 1.01(2) & 3.60(9) &  0.43(6)  & 1.17(3) &  1.63(2) &  1.29(3) &  0.99(2) & 3.49(8)& -0.45(5) & 0.93(6) & 6.58(5) &0.25(1)& 1.06(86) \\ 
     \enddata
    \label{tab:fit_table_rxte}
$^\dagger$ The spectral model is  $wabs*(blackbody + Comptb1 + Comptb2 + Gaussian)$; 
color temperature 
$T_{BB}$ of the bbody component is fixed at 
0.6 keV 
 (see comments in the text); 
$^{\dagger\dagger}$ parameter $\log(A_2)$ is fixed at 2.0 (see comments in the text), 
$^{\dagger\dagger\dagger}$ Normalization parameters of blackbody and Comptb components are in units of 
$L_{37}/d^2_{10}$, where $L_{37}$ is the source luminosity in units of 10$^{37}$ erg/s, 
$d_{10}$ is the distance to the source in units of 10 kpc 
and Gaussian component is in units of $10^{-2}\times total~~photons$ $cm^{-2}s^{-1}$ in line. $^{\dagger\dagger\dagger\dagger}$ $kT_{s1}$ varies  from 1.3 keV to 1.5 keV for  
the {island}  to banana state transition while it is concentrated  about 1.5 keV (see Fig.~\ref{index_temperature_s_12}) for most of the spectra spectra,
$\sigma_{line}$ of Gaussian component is fixed to a value 0.7 keV (see comments in the text),
$N_H$ was free to vary within the range of (2 -- 4)$\times 10^{22}$ cm$^{-2}$ (see comments in the text). 
\end{deluxetable}

\vspace{2.in}
\newpage
~~~~~~~~
\begin{deluxetable}{llccccccc}
\tablewidth{0in}
\tabletypesize{\scriptsize}
    \tablecaption{Comparisons of the Best-fit parameters  of {\it Z}-sources Sco~X-1 and GX~340+0 (1) and  Atoll 
Sources GX~3+1 (2), 4U~1728-34 (3), 4U~1820-30 (4), 4U~1705-44 and  ``Atoll+{\it Z}'' Source XTE~J1701-462 (5)}
    \renewcommand{\arraystretch}{1.2}
 \tablehead
{Source & Alternative & Class$^6$& Distance, & Presence of & $kT_e$,  & $ N_{comptb}$ &  $kT_{s}$  & $f$ \\
  Name  & Name        &          & (kpc)       & kHz QPO     & (keV)         &  $L_{39}^{soft}/{D^2_{10}}$        & (keV)  &  }
 \startdata
Sco~X-1     & V818 Sco & Z, Sp, B     & 2.8 (7) &    +(8)        & 3-180 &  0.3-3.4 & 0.4-1.8 & 0.08-1   \\
4U~1642-45  & GX 340+0 & Z, Sp, B     & 10.5 (9) &    +(12)        & 3-21   &  0.08-0.2  & 1.1-1.5 & 0.01-0.5   \\
4U~1744-26  & GX 3+1   & Atoll, Sp, B & 4.5(10) &     none(13)      & 2.3-4.5 &  0.04-0.15 & 1.16-1.7 & 0.2-0.9   \\
4U~1728-34  & GX~354-0 & Atoll, Su, D & 4.2-6.4(11) & +(14) &  2.5-15& 0.02-0.09 &  1.3 & 0.5-1\\     
4U~1820-30  &   ...    & Atoll, Su, B & 5.8-8 (15) & +(16) &  2.9-21& 0.02-0.14 &  1.1-1.7 & 0.2-1\\     
XTE~J1701-462&  ...    & Atoll+Z, Su, D & 8.8(17) & +(18) &  ...   & ...       &  1-2.7 & ...\\     
4U~1705-44  &   ...    & Atoll, Sp, B & 7.4 (19)  &    +(20) & 2.7-100 &  0.01-0.08 & 1.1-1.5 & 0.2-1   \\
      \enddata
    \label{tab:fit_table_comb}
References:
(1) STF13;
(2) ST12; 
(3) ST11;  
(4) TSF13; 
(5) LRH109;
(6) Classification of the system in the various schemes (see text): Sp = supercritical, Su = subcritical, 
B = bulge, D = disk; 
(7) Bradshaw et al. (1999); 
(8) Zhang et al. (2006); 
(9) Fender \& Henry  (2000), Ford et al. (1998),  Christian \& Swank (1997); 
(10) \citet{kk00}, Ford et al. (2000);  
(11) \citet{par78};  
(12) \citet{Jonker98};  
(13) \citet{stroh98}; 
(14) \citet{to99};  
(15) \citet{ST04};  
(16) \citet{Smale97}; 
(17) \citet{Lin07a}, \citet{lin09}; 
(18) \citet{Sana10}; 
(19) \citet{Haberl_Titarchuk95};  
(20) BO02
\end{deluxetable}

\newpage

\vspace{10.in}

%
%

\newpage

%
%

\newpage

%
%

\newpage

%
%

\newpage 

%
%

\newpage
%
%

\newpage

%
%

\newpage

%
%

\newpage

%
%

\newpage

%
%

\begin{thebibliography}{}

\bibitem[Barret \& Olive (2002)]{barret02}
Barret, D., Olive, J.-F. 2002, ApJ, 578, 391 (BO02)

\bibitem[Boella et al. (1997)]{boel97} 
Boella G., Butler R. C., Perola G. C., et al., 1997, A\&A, 122, 299

\bibitem[Bradshaw et al. (1999)]{brad97} 
Bradshaw, C.F., Fomalont, E.B. \& Geldzahler, B. J, 1999, \apj,  512,  L121

\bibitem[Church  et al. (2014)]{Church14}
Church, M. J., Gibiec, A. \&  Ba\`luci\`nska-Church, M. 2014 MNRAS, 438, 2784

\bibitem[Christian \& Swank (1997)]{cs97} 
Christian, D.J. \& Swank, J. H. 1997, ApJS, 109, 177

\bibitem[D'Ai et al. (2012)]{D'Ai12} 
 D'Ai, A. et al. 2012, A\&A 543, A20 

\bibitem[D'Ai et al. (2010)]{D'Ai10} 
D'Ai, A., Di Salvo, T., Ballantyne, D., et al. 2010, A\&A, 516, id.A36

\bibitem[D'Ai et al. (2007)]{D'Ai07} 
{D'Ai, A., Zycki, P., Di Salvo, T., Iaria, R., Lavagetto, G. \& Robba, N. R.
2007, \apj, 667, 411}

\bibitem[D'Amico et al. (2001)]{D'Amico01}
D'Amico, F., Heindl, W. A., Rothschild, R. E. et al. 2001, ApJ, 547, L147

\bibitem[Di Salvo et al. (2015)]{diSalvo15}
Di Salvo,~T., Iaria, R., A., Iaria, R. et al. 2015, MNRAS, 449, 2794 

\bibitem[Di Salvo et al. (2001)]{diSalvo09}
Di Salvo,~T., D'Ai, A., Iaria, R. et al. 2009, MNRAS, 398, 2022

\bibitem[Di Salvo et al. (2006)]{diSalvo06}
Di Salvo, T., Goldoni, P., Stella, L. et al. 2006, ApJ, 649, L91

\bibitem[Di Salvo et al. (2005)]{disalvo2005}
Di Salvo, T., Iaria, R., Mendez, M., et al. 2005, ApJ, 623, L121

\bibitem[Di Salvo et al. (2003)]{diSalvo03}
Di Salvo, T., Mendez, M. \& van der Klis, M. 2003, A\&A, 406, 177 

\bibitem[Di Salvo et al. (2001)]{disalvo2001}
Di Salvo,~T., Mendez,~M., van der Klis,~M., Ford,~E. \& Robba,~N.R. 2001 \apj, 546, 1107

\bibitem[Di Salvo et al. (2000)]{diSalvo00}
Di Salvo, T., Stella, L., Robba, N. R. et al. 2000, ApJ, 544, L119

\bibitem[Egron et al.  (2013)]{egron13}
Egron, E., Di Salvo, T., Motta, et al. 2013, A\&A, 550, 5

\bibitem[Farinelli et al. (2009)]{Farinelli09}
Farinelli, R., Paizis, A., Landi, R., \& Titarchuk, L. 2009, A\&A, 498, 509

\bibitem[Fender et al. (2000)]{Fender00}
Fender, R. P., \& Hendry, M. A., 2000, MNRAS, 317, 1

\bibitem[Fiocchi et al.  (2007)]{Fiocchi07}
Fiocchi, M., Bazzano, A. \& Ubertini, P.  et al. 2007,  ApJ, 657, 448 

\bibitem[Ford et al. (2000)]{Ford00} 
Ford, E.C. et al. 2000,  \apj,  537, 368  

\bibitem[Ford et al. (1998)]{Ford98}
Ford, E. C., van der Klis, M., \& Kaaret, P. 1998, ApJ, 498, L41

\bibitem[Frontera et al. (1997)]{fron97}
Frontera, F., Costa, E., dal Fiume, D.  et al.  1997, A\&AS, 122, 357

\bibitem[Gierl\`inski \& Done (2002)]{gd02} 
Gierl\`inski, M. \& Done, C. 2002, MNRAS, 331, L47

\bibitem[Haberl \& Titarchuk (1995)]{Haberl_Titarchuk95}
Haberl F. \& Titarchuk L. 1995, A\&A, 299, 414

\bibitem[Hasinger  et al. (1989)]{hasinger89}
Hasinger, G., \& van der Klis, M. 1989, A\&A, 225, 79

\bibitem[Hjellming  et al. (1990)]{Hjellming90}
Hjellming, R. M., Stewart, R. T., White, G. L. 1990, ApJ, 365, 681 


\bibitem[Homan et al. (2010))]{Homan10}
Homan, J., van der Klis, M., Fridriksson, J.~K., 2010, ApJ, 719, 201  

\bibitem[Jonker et al. (1998))]{Jonker98}
Jonker, P. G.,  Wijnands, R., van der Klis, M.  et al. 
1998, ApJ, 499, L191

\bibitem[Kuulkers \& van der Klis  (2000)]{kk00}
Kuulkers, E  \& van der Klis, M. 2000, A\&A, 356, L45

\bibitem[Langmeier et al. (1987))]{Langmeier87}
Langmeier, A. A, Sztajno, M., Hasinger, G. et al. 1987, ApJ, 323, 902

\bibitem[Levine et al. (1996))]{Levine96}
Levine, A. M., Bradt, H., Cui, W., et al. 1996, ApJ, 469, L33


\bibitem[Lin et al.  (2009)]{lin09}
Lin, D.,   Remillard, R.A. \&  Homan, J. 2009, ApJ, 696, 1257  (LRH09)

\bibitem[Lin et al. (2007)]{Lin07a}	
Lin, D., Homan, J., Remillard, R., \& Wijnands, R. 2007, 
Astron. Tel., 1183

\bibitem[Manzo et al. (1997)]{Manzo97}	
Manzo G., Giarrusso S., Santangelo A., et al. 1997, A\&AS 122, 341

\bibitem[Muno et al. (2002)]{muno02}	
Muno, M.P., Remillard, R. A. \& Chakrabarty, D. 2002, ApJ, 568, L35

\bibitem[Piraino et al. (2007)]{Piraino07}
Piraino, S., Santangelo, A., Di Salvo, T., et al. 2007, A\&A 471, L17

\bibitem[Paizis et al.  (2006)]{Paizis06}
Paizis, A., Farinelli, R., Titarchuk, L. et al. 2006, A\&A, 459, 187

\bibitem[Parmar et al.  (1997)]{parmar97}
Parmar, A., Martin, D. D. E., Bavdaz, M. et al. 1997, A\&AS, 122, 309


\bibitem[Sanna  et al. (2010)]{Sana10}
Sanna, A., M$\acute e$ndez, M., Altamirano, D., Homan, J., et al. 2010, MNRAS, 408, 622 

\bibitem[Sztajno et al.  (1985)]{Sztajno85}
Sztajno M., Langmeier, A., Frank, J. et al. 1985, IAU Circ No. 4111

\bibitem[Seiﬁna et al. (2014)]{STS14} 
Seifina, E., Titarchuk, L., \& Shaposhnikov, N. 2013, ApJ, 789, 57 (STS14)

\bibitem[Seiﬁna et al. (2013)]{STF13} 
Seifina, E., Titarchuk, L., \& Frontera, F. 2013, ApJ, 766, 63 (STF13)

\bibitem[Seifina \& Titarchuk  (2012)]{ST12} Seifina, E. \& Titarchuk, L. 2012,
 \apj, 747, 99 (ST12)

\bibitem[Seifina \& Titarchuk  (2011)]{ST11} Seifina, E. \& Titarchuk, L. 2011,
 \apj, 738, 128 (ST11)


\bibitem[Shakura \& Sunyaev  (1973)]{ss73} Shakura, N.I., \& Sunyaev, R.A. 1973, \aap, 24, 337  


\bibitem[Shaposhnikov \& Titarchuk  (2004)]{ST04} Shaposhnikov, N. \&  Titarchuk  L.  2004, \apj, 606, L57

\bibitem[Smale et al. (1997)]{Smale97}
Smale, A. P., Zhang, W., \& White, N. E. 1997, ApJL, 483, L119

\bibitem[Strohmayer (1998)]{stroh98}
 Strohmayer, T.  1998, AIP Conference Proceedings, Volume 431, pp. 397-400  in American Institute of Physics Conference Series, (astro-ph/9802022v1)

\bibitem[Sunyaev \& Titarchuk  (1980)]{st80}  
Sunyaev, R.A. \& Titarchuk, L. 1980, A\&A, 86, 121

\bibitem[Titarchuk et al. (2014)]{TSS14}  
Titarchuk, L., Seifina, E. \& Shrader, Ch. 2014, ApJ, 798, 98 (TSS14)


\bibitem[Titarchuk et al. (2013)]{TSF13}  
Titarchuk, L., Seifina, E., \& Frontera, F. 2013, ApJ, 767, 160 (TSF13)

\bibitem[Titarchuk \& Osherovich (1999)]{to99} 
Titarchuk, L.G. \& Osherovich, V.A. 1999, \apj, 518, L95

\bibitem[Titarchuk et al. (1998)]{tlm98} 
Titarchuk, L., Lapidus, I.I.  \& Muslimov, A. 1998, \apj, 419, 315 


\bibitem[van Paradijs  (1978)]{par78} 
van Paradijs, J. 1978,  Nature, 274, 650

\bibitem[van Straaten et al.  (2000)]{vStraat00}
van Straaten, S., Ford, E., van der Klis, M., M$\acute e$ndez, M. \& 
Kaaret, Ph. 2000, ApJ, 540, 1049

\bibitem[Zhang et al.  (2006)]{zhang06}
Zhang, C.M. et al. 2006, MNRAS, 366, 1373

\end{thebibliography}
\end{document}